\documentclass[12pt]{iopart}
\usepackage{graphicx}
\usepackage{iopams}
\usepackage{setstack}
\usepackage{textcomp}

\usepackage{color}
\usepackage{psfrag}

\usepackage{epsfig}

\begin{document}

\title[High-dimensional kagom\'e and diamond crystals]{High-dimensional generalizations of the
kagom\'e and diamond crystals and the decorrelation principle for periodic sphere packings}

\author{Chase E.~Zachary$^1$ and Salvatore Torquato$^2$}
\address{$^1$ Department of Chemistry, Princeton University, Princeton, New Jersey 08544, USA}
\address{$^2$ Department of Chemistry, Department of Physics, Princeton Center for 
Theoretical Science, Program in Applied and Computational Mathematics, and 
Princeton Institute for the Science and Technology of Materials, 
Princeton University, Princeton, New Jersey 08544, USA}
\ead{torquato@electron.princeton.edu}

\begin{abstract}
In this paper, we introduce constructions of the high-dimensional generalizations of the 
kagom\'e and diamond crystals.
The two-dimensional kagom\'e crystal and its three-dimensional counterpart, the 
pyrochlore crystal, have been extensively studied in the context of geometric frustration
in antiferromagnetic materials.  Similarly, the polymorphs of elemental carbon 
include the diamond crystal and the corresponding two-dimensional honeycomb structure,
adopted by graphene.
The kagom\'e crystal in $d$ Euclidean dimensions consists of 
vertex-sharing $d$-dimensional simplices in which all of the points are topologically equivalent.
The $d$-dimensional generalization of the diamond crystal can then be obtained from the centroids 
of each of the simplices, and we show that this natural construction of the diamond crystal is 
distinct from the $D_d^+$ family of crystals for all dimensions $d\neq 3$.   We analyze the 
structural properties of these high-dimensional crystals, including the packing densities, coordination
numbers, void exclusion probability functions, covering radii, and quantizer errors.   Our results
demonstrate that the so-called decorrelation principle, which formally
states that unconstrained correlations vanish in asymptotically high dimensions, 
remarkably applies to 
the case of periodic point patterns with inherent long-range order.  We argue that
the decorrelation principle is already exhibited 
in periodic crystals in low dimensions via  a ``smoothed'' pair correlation function 
obtained by convolution with a Gaussian kernel.  These observations support the 
universality of the decorrelation principle for any point pattern in high dimensions, whether
disordered or not. 
This universal property in turn 
suggests that the best conjectural lower bound on the maximal 
sphere-packing density in high Euclidean dimensions
derived by Torquato and Stillinger [Expt. Math. \textbf{15}, 307 (2006)] is in fact optimal.
\end{abstract}

\maketitle

\section{Introduction}

There has been substantial recent interest in the physics of high-dimensional systems
with applications to ground-state problems, packing problems, number theory, and 
phase behaviors of many-particle systems \cite{To10, ToSt03, ZaTo09, Sk06, Pa06, Ro07, Me09, Lu10}.  
The problem of identifying the densest sphere packings in high-dimensional Euclidean space 
$\mathbb{R}^d$ is an open and fundamental problem in discrete geometry and number theory 
with important applications to communications theory \cite{CoSl99}.  In particular, Shannon
showed that the optimal method of sending digital signals over noisy channels corresponds to the
densest packing in a high-dimensional space \cite{Sh48}.  
Although the densest packings in dimensions
two and three are known to be Bravais lattice packings (the triangular and FCC lattices, respectively
\cite{Ha05}), in sufficiently high dimensions, non-Bravais lattices
almost surely are the densest packings. In addition to 
providing putative exponential improvement on Minkowski's lower bound for the maximal 
sphere-packing density, Torquato and Stillinger presented strong arguments to suggest that the 
densest packings in high dimensions are in fact \emph{disordered} \cite{ToSt06, ToSt10}.  

Their methods rely on the so-called \emph{decorrelation principle} for disordered sphere packings,
which states that as the dimension $d$ increases, all \emph{unconstrained} correlations 
asymptotically vanish, and any higher-order correlation functions $g_n(\mathbf{r}^n)$  
may be expressed in terms of the number density $\rho$ and the pair correlation function 
$g_2$ \cite{ToSt06}.  Since its introduction, additional work has shown that the decorrelation
principle is remarkably robust, meaning that it 
is already manifested in low dimensions and applies also to certain soft-matter systems \cite{Za08} and 
quantum many-particle distributions \cite{To08}.  Furthermore, detailed numerical studies of saturated
maximally random jammed hard-sphere packings, which are 
the most disordered packings of spheres that are 
rigorously incompressible and nonshearable, have demonstrated that 
unconstrained correlations beyond the hard-sphere diameter asymptotically vanish even in 
relatively low dimensions $d = 1$-$6$ \cite{Sk06}.   Similar results have also been observed 
for the exactly-solvable ``ghost'' random sequential addition (RSA) 
process \cite{To06B} along with the usual RSA process \cite{ToUcSt06}.  All evidence
to date supports the notion that the decorrelation principle applies fundamentally
to disordered many-particle systems.  In this paper,
we provide evidence that the decorrelation principle applies more generally to any periodic 
crystal, which has implications for the densest sphere packings
in high dimensions and sheds light on the reasons why
it is robust in low dimensions.

The properties of periodic crystal structures are fundamental to the physical and mathematical 
sciences.  Experience in two- and three-dimensional systems suggests that crystals are 
prototypical ground states of matter \cite{ToSt08}, obtained by slow annealing of an interacting 
many-particle system to absolute zero temperature.  Unlike disordered states of matter, including
gases and liquids, crystals possess complete long-range correlations and translational symmetry.  
As such, periodic crystals can be specified by translational copies of a single fundamental cell 
containing one (in the case of a Bravais lattice) or more particles.

Elemental carbon is known to adopt numerous polymorphs of fundamental significance.  Its 
four-electron valence structure implies that it can be tetrahedrally, covalently bonded with itself 
to form the diamond crystal.  Certain ``superdense'' 
polymorphs of carbon involving different packings of 
carbon tetrahedra have also recently been reported in the 
literature \cite{Zh11}.   The two-dimensional analog of the diamond crystal is the 
so-called honeycomb crystal, in which points are placed at the vertices of hexagons 
that tile the plane.  This polymorph of carbon is the graphene structure, variations of which have gained
substantial interest as nanomaterials \cite{GeNo07}.  Each point in the honeycomb crystal
is coordinated with three other points of the structure, and by placing particles at the midpoints of 
each of the ``bonds,'' one obtains the kagom\'e crystal.  
The kagom\'e crystal
and its three-dimensional counterpart, the pyrochlore crystal, have 
been used in models of spin-frustrated antiferromagnetic materials \cite{So05}.  
This type of geometric frustration in so-called ``spin ice'' 
induces a nonvanishing residual entropy in the ground state, 
equivalent to behavior identified in liquid water \cite{Pa35}.

Recently, Torquato has reformulated the covering and quantizer problems from 
discrete geometry as ground-state problems involving many-body interactions with 
one-body, two-body, three-body, and higher-body potentials \cite{To10}.  
Formally, the covering problem seeks the point configuration that minimizes 
the radius of overlapping spheres circumscribed around each of the points
required to cover $\mathbb{R}^d$ \cite{CoSl99, To10}.  The quantizer problem involves finding the 
point configuration in $\mathbb{R}^d$ that minimizes a Euclidean ``distance error''
associated with replacing a randomly placed point in $\mathbb{R}^d$ 
with the nearest point of the point process \cite{CoSl99, To10}.
Closely related 
is the so-called number variance problem, which aims to identify the distribution of points 
that minimizes fluctuations in the local number density over large length scales \cite{ToSt03, 
ZaTo09}.  This problem can also be interpreted as the determination of the ground state
for a particular soft, bounded pair interaction, and, for the special case of Bravais lattices, 
is equivalent to identifying the minimizer of the so-called Epstein zeta function \cite{Sa06}.
Note that the number variance of a point pattern has been suggested to quantify structural order 
over large length scales \cite{ToSt03, ZaTo09}.

Studies of many-body fluids and amorphous packings have attempted to glean new information about 
low-dimensional physical properties, including the equation of state,
radius of convergence of the virial series, phase transitions, 
and structure, from high-dimensional models.   Frisch and Percus have 
shown that for repulsive interactions, Mayer cluster expansions of the free energy
are dominated by ring diagrams at each order in particle density $\rho$ \cite{Fr99}.  
This result was extended by Zachary, Stillinger, and Torquato to show that the 
so-called mean-field approximation for soft, bounded pair interactions becomes exact
in the high-dimensional limit \cite{Za08}.   Parisi and Zamponi have utilized 
the HNC approximation to the pair correlation function and mean-field theory to 
understand hard-sphere glasses and jamming in high dimensions \cite{Pa06}, and Rohrmann 
and Santos have generalized results from liquid-state theory to study fluids
of hard spheres in high dimensions \cite{Ro07}.   Michels and Trappeniers
\cite{Mi84}, Skoge {\it et al.} \cite{Sk06}, 
van Meel {\it et al.} \cite{Me09}, and Lue \emph{et al.} \cite{Lu10} have numerically studied the effect
of dimensionality of the disorder-order transition 
in equilibrium hard-sphere systems in up to dimension six.  Additionally, Doren and Herschbach 
have developed a dimensionally-dependent perturbation theory for quantum systems
to draw conclusions about the energy eigenvalues in low dimensions \cite{Do86}.  

In this paper, we generalize the kagom\'e and diamond crystals for 
high-dimensional Euclidean space $\mathbb{R}^d$.  We are motivated by the observation that there
are $d+1$ particles within the fundamental cell of the kagom\'e crystal, which grows with the 
dimension. The high-dimensional kagom\'e crystal thus possesses a large basis 
of particles and approximates the case of a (possibly irregular) 
$N$-particle many-particle distribution subject to 
periodic boundary conditions for $N$ large.  The $d$-dimensional kagom\'e crystal  
provides an intriguing structure for which to test the applicability of the decorrelation 
principle for periodic point patterns.  Since such periodic crystals possess full long-range order, 
it is highly nonintuitive that the decorrelation principle should apply, and yet we provide  
indirect and direct evidence that it continues to hold in this general setting.  Furthermore, 
by analyzing the structural properties of the high-dimensional diamond and kagom\'e crystals, 
we show that certain ``disordered'' packings can be quantitatively more ordered with respect to 
local fluctuations in the number density than periodic crystals, even in relatively low dimensions.  
Our results therefore have important implications for the low- and high-dimensional problems outlined
above.




Our major results are summarized as follows:

\begin{enumerate}
\item  We develop constructions of high-dimensional generalizations of the kagom\'e and 
diamond crystals using the geometry of the fundamental cell for the $A_d$ Bravais lattice
(defined below).  Our results suggest a natural method for constructing a large class of 
``kagom\'e-like'' crystals in high dimensions.
\item  We examine the behavior of structural features of the kagom\'e and diamond crystals, 
including the packing densities, coordination numbers, covering radii, and quantizer errors.  
In particular, we show that the kagom\'e crystal possesses a lower packing fraction than the 
diamond crystal for all $d \geq 4$, a larger covering radius for all $d \geq 2$, and a larger
quantizer error for all $d \geq 3$.
\item  We relate these structural features to the distribution of the void space external to the 
particles in the fundamental cell via numerical calculation of the void exclusion probability function 
$E_V$ (defined below).  As the spatial dimension increases, the fundamental cell of the kagom\'e
lattice develops substantially large holes, thereby skewing the bulk of the void-space distribution 
such that large holes are less rare than in the uncorrelated Poisson point pattern.  The kagom\'e 
crystal therefore lies above Zador's upper bound on the minimal quantizer error in sufficiently high 
dimensions.
\item  We calculate the number variance coefficients governing asymptotic surface-area
fluctuations in the local number density for the kagom\'e and diamond crystals.  
The kagom\'e crystal for all $d \geq 3$ possesses a larger number variance coefficient than 
a certain correlated disordered packing corresponding to a so-called $g_2$-invariant process 
\cite{ToSt06, St05, ToSt02}, providing indirect evidence for a decorrelation principle of high-dimensional 
periodic point patterns.
\item  We provide direct evidence for a decorrelation principle of periodic structures by 
examining a ``smoothed'' pair correlation function for the $d$-dimensional kagom\'e crystal.
Our analysis also applies to Bravais lattices as evidenced by corresponding calculations 
for the hypercubic lattice $\mathbb{Z}^d$, establishing the universality of the decorrelation effect.  
These results suggest that pair correlations
alone are sufficient to completely characterize a sphere packing for large 
dimension $d$ and that the best conjectural lower bound on the maximal density of 
sphere packings provided by Torquato and Stillinger \cite{ToSt06} may in fact be optimal in 
high dimensions.    This statement suggests that the densest sphere packings in high 
dimensions are in fact disordered.  
\end{enumerate}

\section{Definitions}

\subsection{Crystals and correlation functions}

A $d$-dimensional \emph{Bravais lattice} is a periodic structure defined by integer 
linear combinations of a set of basis vectors $\{\mathbf{e}_j\}$ for $\mathbb{R}^d$, i.e., 
\begin{eqnarray}
\mathbf{p} &= \sum_{j=1}^d n_j \mathbf{e}_j \equiv M_{\Lambda}\mathbf{n} 
\qquad (n_j \in \mathbb{Z} ~~\forall j)
\end{eqnarray}
for all points $\mathbf{p}$ of the Bravais lattice \cite{CoSl99}, where we have
defined the generator matrix $M_{\Lambda}$ of the Bravais lattice $\Lambda$ with columns given 
by the basis vectors.  The basis vectors define 
a \emph{fundamental cell} for the Bravais lattice containing only one lattice point.  This concept 
can be naturally generalized to include multiple points within the fundamental cell, defining a 
periodic crystal, or
non-Bravais lattice \cite{FN1}. 
Specifically, a non-Bravais lattice consists of the union of a Bravais lattice with one or more
translates of itself; it can therefore be defined by specifying the generator matrix $M_{\Lambda}$ for the 
Bravais lattice along with a set of translate vectors $\{\boldsymbol\nu_j\}$.   Note that the special case of 
a single zero translate vector $\mathbf{0}$ defines a Bravais lattice.  

Every Bravais lattice $\Lambda$ possesses a \emph{dual Bravais lattice} 
$\Lambda^*$ with lattice points 
$\mathbf{q}$ defined by $\mathbf{p}\cdot\mathbf{q} = m \in \mathbb{Z}$ \cite{FN2}.  The generator
matrix for the dual Bravais lattice is given by \cite{CoSl99}
\begin{eqnarray}
M_{\Lambda^*} &= (M_{\Lambda}^{\mbox{\scriptsize T}})^{-1},
\end{eqnarray}
where $B^{\mbox{\scriptsize T}}$ denotes the transpose of a matrix $B$.  A Bravais lattice and its dual obey 
the Poisson summation formula \cite{FNPS} for any Schwartz function \cite{ToSt08, CoKuSc09}.  
In general, a crystal, containing more than one particle per fundamental cell, 
does not possess a dual structure in the same sense as for a Bravais lattice.  

A many-particle distribution is determined by its number density $\rho$, equal to the number 
of particles per unit volume, and 
the set of \emph{$n$-particle correlation functions} $g_n$, proportional to the probability 
density of finding a configuration $\mathbf{r}^n$ of any $n$ particles within the system.  
Of particular importance is the pair correlation function 
$g_2(r)$, which for an isotropic and statistically homogeneous point pattern is a function only of the 
magnitude $r$ of pair separations between particles.  For any periodic crystal consisting of 
topologically equivalent particles, the angularly-averaged pair correlation function has the form
\begin{eqnarray}
\rho s(r)g_2(r) &= \sum_{k=1}^{+\infty} Z_k \delta(r-r_k),\label{g2periodic}
\end{eqnarray}
where $Z_k$ is the number of points at a radial distance $r_k$ away from a reference particle of the 
lattice and $s(r)$ is the surface area of a $d$-dimensional sphere of radius $r$.  The 
\emph{cumulative coordination number} $Z(R)$, the total number of particles within a radial distance 
$R$ from a reference particle, is therefore given by
\begin{eqnarray}
Z(R) &= \rho\int_0^{R} s(r) g_2(r) dr;\label{ZR}
\end{eqnarray}
for a periodic crystal, this identity simplifies to $Z(R) = \sum_{k=1}^K Z_k$, where $K$
denotes the highest index for which $r_K \leq R$.  

\subsection{Hyperuniformity and the number variance problem}

Torquato and Stillinger have characterized fluctuations in the local number density of 
a many-particle distribution \cite{ToSt03} and have shown that these fluctuations
behave differently for periodic crystals and uncorrelated systems.
Define the random variable $N(\mathbf{x}_0; R)$ to the be the number of particles 
within a spherical observation window of radius $R$ centered at position $\mathbf{x}_0$.  
By definition, $\langle N(\mathbf{x}_0; R)\rangle = \rho v(R)$, where $v(R)$ is the volume of a 
$d$-dimensional sphere of radius $R$.  For a Poisson point pattern in which there are no 
correlations between particles, the underlying Poisson counting measure also implies that
\begin{eqnarray}
\sigma^2(R) &= \langle N^2(\mathbf{x}_0; R)\rangle - \langle N(\mathbf{x}_0; R)\rangle^2
= \langle N(\mathbf{x}_0; R)\rangle = \rho v(R), 
\end{eqnarray}
meaning that fluctuations in the local number density of the observation window scale with 
the window volume.  

However, this scaling is not a general feature of all point patterns.  
In the general case of correlated point patterns, the local number variance is
given by \cite{ToSt03}
\begin{eqnarray}
\sigma^2(R) &= \rho v(R)\left\{1+\rho\int \left[g_2(\mathbf{r})-1\right]\alpha(r; R) d\mathbf{r}\right\},
\end{eqnarray}
where $\alpha(r; R)$ is the so-called \emph{scaled intersection volume}, defined geometrically
as the volume of the intersection of two $d$-dimensional spheres of radius $R$ with centers separated
by a distance $r$, normalized by the volume $v(R)$ of a sphere.  Explicit expressions for 
the scaled intersection volume in various dimensions have been given by Torquato and Stillinger 
\cite{ToSt03, ToSt06}.  

Introducing a length scale $D$ (e.g., the mean nearest-neighbor distance between points)
and a corresponding reduced density $\eta = \rho v(D/2)$,
one can show that the asymptotic behavior of the number variance for large observation windows is 
\cite{ToSt03}
\begin{eqnarray}
\sigma^2(R) &= 2^d \eta\left\{ A \left(R/D\right)^d + B \left(R/D\right)^{d-1} + 
o[(R/D)^{d-1}]\right\},
\end{eqnarray}
where $o(x)$ denotes terms of order less than $x$.  The coefficients $A$ and $B$ 
are given by 
\begin{eqnarray}
A &= 1+\rho\int [g_2(\mathbf{r})-1] d\mathbf{r} = \lim_{\vert\vert\mathbf{k}\vert\vert\rightarrow 0} S(\mathbf{k})\\
B &= \frac{-\eta \Gamma(1+d/2)}{D v(D/2) \Gamma[(d+1)/2] \Gamma(1/2)} 
\int \vert\vert\mathbf{r}\vert\vert [g_2(\mathbf{r})-1] d\mathbf{r}\label{BNV},
\end{eqnarray}
where $S(\mathbf{k}) = 1+\rho\mathfrak{F}\{g_2(\mathbf{r})-1\}(\mathbf{k})$, with $\mathfrak{F}$
denoting the Fourier transform, is the 
\emph{structure factor}.
It follows that the number variance of any point pattern for which $A = 0$ grows more 
slowly than the volume of the observation window, implying that the point pattern 
is \emph{effectively} homogeneous even on local length scales.  Such systems are 
known as \emph{hyperuniform} \cite{ToSt03} or \emph{superhomogeneous} point patterns \cite{PiGaLa02}.  
Examples of hyperuniform point patterns include all Bravais and non-Bravais lattices, quasicrystals
possessing Bragg peaks, and certain disordered point patterns with pair correlations decaying 
exponentially fast.  

It has also been suggested that the coefficient $B$ quantifies 
large-scale order in a hyperuniform point pattern \cite{ToSt03, ZaTo09}.  The issue of identifying the 
point pattern that minimizes this coefficient defines the so-called number variance problem \cite{To10, ToSt03}.
It has recently been proved that the integer lattice is the unique number variance minimizer in 
one dimension among all hyperuniform point patterns \cite{ToSt03}.  
Numerical results strongly suggest that the 
triangular lattice minimizes the number variance in two dimensions \cite{ToSt03, ZaTo09}.  
However, contrary to the 
expectation that the densest lattice packing should also minimize the number variance, it has been
shown in three dimensions that the BCC lattice possesses a lower asymptotic number variance 
coefficient $B$ than the FCC lattice \cite{ToSt03}.  
It is worth mentioning in this regard that the BCC lattice is the 
dual of FCC.

\subsection{Jamming in hard-sphere packings}

A \emph{sphere packing} is obtained from a point pattern in $d$-dimensional Euclidean space
by decorating each of the points with a sphere of radius $R_P$ such that no spheres overlap 
after the decoration; the parameter $R_P$ is the packing radius.    It is an open and nontrivial 
problem to quantify the extent of randomness (equivalently, of order) in a sphere packing, 
which reflects nontrivial structural information about the system.  Research in this area is aimed
at identifying sets of \emph{order metrics} $\psi$ \cite{ToTrDe00} that align with physical intuitions of order,
at least in relatively low dimensions, and are positively correlated.  It has recently been proposed that 
hyperuniformity is itself a measure of order over large length scales \cite{ToSt03, ZaTo09}.

Torquato and Stillinger have introduced a classification of sphere packings in terms of the extent to 
which they are \emph{jammed} \cite{ToSt01, ToSt10}.  
In particular, they have provided a mathematically precise 
hierarchy of jammed sphere packings, distinguished depending on the nature of their 
mechanical stability \cite{ToSt01, ToSt10}:
\begin{enumerate}
\item  \emph{Local jamming}:  Each particle in the packing is locally trapped by at least
$d+1$ contacting neighbors, not all in the same hemisphere.  Locally jammed 
particles cannot be translated while fixing the positions of all other particles.
\item \emph{Collective jamming}:  Any locally jammed configuration is collectively jammed 
if no subset of particles can simultaneously be displaced so that its members move out of contact 
with each other and with the remainder set.
\item  \emph{Strict jamming}:  Any collectively jammed configuration is strictly jammed if it 
disallows all uniform volume-nonincreasing strains of the system boundary.
\end{enumerate}
These categories certainly do not include all possible distinctions of jammed configurations, but 
they span a reasonable spectrum of possibilities.   Importantly, jamming depends explicitly 
on the boundary conditions for the packing \cite{ToSt01, ToSt10}.

\emph{Isostatic} packings are jammed packings with the minimal number of contacts $M$ for 
a given jamming category under the specified boundary conditions \cite{ToSt10}.  Under periodic 
boundary conditions, for collective jamming $M = 2N-1$ and $3N-2$ for $d = 2$ and $d = 3$, 
respectively, and for strict jamming $M = 2N+1$ and $3N+3$ for $d = 2$ and $d = 3$, respectively 
\cite{ToSt10, DoCoStTo07}.  
In this case, the relative differences between isostatic collective and strictly jammed packings
diminishes for large $N$, and an isostatic packing
in $d$ dimensions has a mean contact number per particle $Z = 2d$,
known as the isostatic condition.  Note, however, that
packings for which $Z = 2d$ are not necessarily collectively or strictly jammed; 
the two-dimensional square lattice and three-dimensional simple-cubic lattice are
simple counterexamples in dimensions $d = 2$ and $d = 3$, respectively.  Another interesting 
example is the two-dimensional kagom\'e crystal, which is locally jammed but neither collectively nor 
strictly jammed under periodic 
boundary conditions and possesses a nearest-neighbor contact number per particle $Z = 4$ 
\cite{ToSt01, DoCoStTo07}.  However, this
structure can be made strictly jammed by ``reinforcing'' it with an extra ``row'' and
``column'' of disks \cite{DoCoStTo07}.

\subsection{The covering problem}

Consider a distribution of particles at unit number density.  The \emph{covering radius} for 
the point process is defined by decorating each of the particles with a sphere of radius $R$
and identifying the minimal radius $R_C$ necessary to cover the space completely.  More precisely,
for any choice of $R$, we can define the volume fraction of space $\phi_P$ occupied by the
spheres; the volume fraction occupied by the void space external to the spheres is then 
$\phi_V = 1-\phi_P$.  The covering radius $R_C$ is then defined as the 
minimal value of $R$ for which $\phi_P = 1$ and $\phi_V = 0$.  

The volume fraction $\phi_V$ of the void space external to a set of spheres of radius $R$ is 
equivalent to the probability of inserting a ``test'' sphere of radius $R$ into the system and finding it 
contained entirely in the void space.  This latter quantity is known as the \emph{void exclusion 
probability function} $E_V(R)$
and can be expressed in terms of the $n$-particle correlation functions
for the underlying point pattern \cite{ToLuRu90, To02}:
\begin{eqnarray}
E_V(R) &= 1+\sum_{k=1}^{+\infty} \frac{(-\rho)^k}{\Gamma(k+1)} \int
g_k(\mathbf{r}^k) \prod_{j=1}^k \Theta(R-\vert\vert \mathbf{x}-\mathbf{r}_j\vert\vert) d\mathbf{r}_j,
\end{eqnarray}
where $\Theta(x)$ is the Heaviside step function.  This expression can also be rewritten
for a statistically homogeneous point pattern in terms of 
intersection volumes of spheres \cite{To10}:
\begin{eqnarray}
E_V(R) &= 1+\sum_{k=1}^{+\infty} \frac{(-\rho)^k}{\Gamma(k+1)} 
\int g_k(\mathbf{r}_{12}, \ldots, \mathbf{r}_{1k})
v_{\mbox{\scriptsize int}}^{(k)} (\mathbf{r}^k; R) d\mathbf{r}^k,\label{Evvint}
\end{eqnarray}
where $v_{\mbox{\scriptsize int}}^{(k)}(\mathbf{r}^k; R)$ is the intersection volume of $k$ spheres of radius $R$
and centers at $\mathbf{r}^k$:
\begin{eqnarray}
v_{\mbox{\scriptsize int}}^{(k)}(\mathbf{r}^k; R) &= \int d\mathbf{x} \prod_{j=1}^k 
\Theta(R-\vert\vert\mathbf{x}-\mathbf{r}_j
\vert\vert).
\end{eqnarray}
The expression \eref{Evvint} implies that $E_V(R)$ can be interpreted as a total energy 
per particle associated with a many-particle interaction involving one-body, two-body, 
three-body, and higher-body potential energies \cite{To10}.  For a single realization of
$N$ points in a volume $V \subset \mathbb{R}^d$ \cite{To10}
\begin{eqnarray}
\fl E_V(R) &= 1-\rho v(R) + \frac{1}{V}\sum_{i<j} v_{\mbox{\scriptsize int}}^{(2)}(r_{ij}; R) 
-\frac{1}{V} \sum_{i<j<k} v_{\mbox{\scriptsize int}}^{(3)}(r_{ij}, r_{ik}, r_{jk}; R)+\cdots.
\end{eqnarray}  
We remark that truncating the series expression \eref{Evvint} at order $k$
provides an upper bound to $E_V$ when $k$ is even and a lower bound when $k$ is odd \cite{To86}.

The covering problem concerns identifying the point pattern with the minimal covering radius
at unit number density.  
In particular, one attempts to identify the 
point pattern that minimizes the one-dimensional Lebesgue 
measure of the interval of compact support $[0, R_C]$ of the 
void exclusion probability function $E_V(R)$.  A lower bound on the minimal covering radius 
can be obtained by truncating the series representation \eref{Evvint} for $E_V$ at first order, 
implying at unit number density that
\begin{eqnarray}
E_V(R) &\geq \left[1- v(R)\right]\Theta\left[1-v(R)\right].
\end{eqnarray}
This lower bound has a zero at $R^* = \Gamma^{1/d}(1+d/2)/\sqrt{\pi}$, which increases as 
$\sqrt{d}$ for large $d$.

\subsection{The quantizer problem}

A $d$-dimensional quantizer is a device that takes as an input a point at position $\mathbf{x}$ in 
$\mathbb{R}^d$ generated from some probability density function $p(\mathbf{x})$ and outputs the 
nearest point $\mathbf{r}_j$ of a known point pattern to $\mathbf{x}$ \cite{CoSl99}.  The quantizer problem 
is then to choose the point pattern to minimize the \emph{scaled dimensionless error}
$\mathcal{G} = \langle R^2\rangle/d$, where $\langle R^2\rangle$ is the second moment of the 
nearest-neighbor distribution function for the \emph{void space} external to the particles in the 
point process.  Specifically, we define the \emph{void nearest-neighbor density function} $H_V(R)$
such that $H_V(R) dR$ is the probability of finding the nearest particle of a point pattern with respect to 
an arbitrary point $\mathbf{x}$ of the void space within a radial distance $R+dR$ from $\mathbf{x}$.

The void exclusion probability function $E_V(R)$ is then the complementary cumulative distribution 
function associated with $H_V(R)$ \cite{To02}:
\begin{eqnarray}
E_V(R) &= 1-\int_0^R H_V(r) dr.
\end{eqnarray}
Using integration by parts, one can then show that
\begin{eqnarray}
\mathcal{G} &= \frac{1}{d} \int_0^{+\infty} R^2 H_V(R) dR\\
&= \frac{2}{d} \int_{0}^{+\infty} R E_V(R) dR. 
\end{eqnarray}
The quantizer error therefore depends sensitively on the shape of the void-space distribution.  This 
situation is distinct from the covering problem, which is concerned only with the compact support of 
$E_V$.    

Using upper and lower bounds on $E_V$, Torquato has been able to re-derive Zador's bounds
for the minimum scaled dimensionless error \cite{To10}:
\begin{eqnarray}
\frac{\Gamma^{2/d}(1+d/2)}{\pi(d+2)} &\leq \mathcal{G}_{\mbox{\scriptsize min}} 
\leq \frac{\Gamma^{2/d}(1+d/2)\Gamma(1+2/d)}{\pi d}.\label{Zador}
\end{eqnarray}
These bounds converge in asymptotically high dimensions, implying
\begin{eqnarray}
\mathcal{G}_{\mbox{\scriptsize min}} &\rightarrow (2\pi e)^{-1} \qquad (d\rightarrow +\infty).\label{Gasymp}
\end{eqnarray}
This convergence implies that lattices and disordered point patterns are equally good 
quantizers in asymptotically high dimensions.  Using known results for sphere packings,
Torquato has also presented an improved upper bound to the minimal quantizer error \cite{To10},
which is generally appreciably tighter than Zador's upper bound for low to moderately high
dimensions and converges to the exact asymptotic result \eref{Gasymp} in high dimensions.


\subsection{Comparison of the packing, number variance, covering, and quantizer problems}

In his study of the best solutions of the covering and quantizer problems in up to 
$24$ dimensions, Torquato \cite{To10} compared these results to the best known solutions 
for the sphere packing and number variance problems.
In $\mathbb{R}$ and $\mathbb{R}^2$, it is well-known that the integer lattice $\mathbb{Z}$ and the 
triangular lattice, respectively, possess simultaneously the maximal packing density, the minimal 
asymptotic number variance, the minimal covering radius, and the 
minimal quantizer error \cite{CoSl99, To10}.  However, the solutions to these problems 
are no longer the same in as low as three dimensions.  Although the FCC lattice generates 
the densest sphere packing in three dimensions \cite{Ha05}, its \emph{dual} lattice BCC
minimizes the three-dimensional covering radius, quantizer error, and asymptotic
number variance.  To understand these differences,
Torquato has shown that while the number variance, covering, and quantizer problems
are described by soft, bounded interactions, the packing problem is described by 
a short-ranged pair potential that is zero whenever two spheres do not overlap and 
infinite when they do \cite{To10}.   Furthermore, although the number variance problem 
can be interpreted as the determination of the ground state of a short-ranged soft pair interaction 
\cite{ToSt03, To10}, the covering and quantizer problems involve one-body, two-body, 
three-body, and higher-order interactions \cite{To10}.  Therefore,  for $d \geq 4$,
the solutions for each of these problems are not necessarily the same.  
One notable exception occurs in $\mathbb{R}^{24}$, where the Leech lattice $\Lambda_{24}$ 
\cite{CoSl99} likely provides the globally optimal solution for all four problems \cite{To10}.  
It is currently unknown whether such globally optimal solutions exist for dimensions other 
than $d = 1$, $d = 2$, and $d = 24$.  It was shown \cite{To10} that \emph{disordered} saturated 
sphere packings provide both good coverings and quantizers in relatively low dimensions
and may even surpass the best known lattice coverings and quantizers in these dimensions.
We shall return to this point in Section IV.

\section{High-dimensional generalizations of the kagom\'e and diamond crystals}


Our constructions of the $d$-dimensional generalizations of the kagom\'e and diamond crystals
will involve an underlying $A_d$ Bravais lattice structure.  All angles between the basis vectors for
the $A_d$ lattice are $\pi/3$ radians, implying that $\mathbf{e}_j\cdot \mathbf{e}_k = a^2/2$ for all 
$j \neq k$, where $a$ is the magnitude of each basis vector $\mathbf{e}_j$ \cite{FNAd}.  It is therefore possible to 
identify a coordinate system in which the generator matrix $M_{A_d}$ is triangular.  The two-dimensional
$A_2$ lattice is the usual triangular lattice, which is the known densest packing
in $\mathbb{R}^2$.  Similarly, the $A_3$ lattice is one representation for the FCC lattice, which 
is the densest packing in three dimensions \cite{Ha05}.  However, for $d \geq 4$, the 
$A_d$ is no longer optimally dense, even among Bravais lattices.

\subsection{The $d$-dimensional diamond crystal}

The fundamental cell for the $A_d$ lattice is a regular rhombotope, the $d$-dimensional 
generalization of the two-dimensional rhombus and three-dimensional rhombohedron.  
Therefore, the points $\{\mathbf{0}\} \cup \{\mathbf{e}_j\}_{j=1}^d$,
where $\mathbf{e}_j$ denotes a basis vector of the $A_d$ lattice, are situated at the 
vertices of a regular $d$-dimensional simplex.  The $d$-dimensional diamond crystal 
can therefore be obtained by including in the fundamental cell the centroid of this 
simplex:
\begin{eqnarray}
\boldsymbol\nu &= \frac{1}{d+1}\sum_{j=1}^d \mathbf{e}_j,\label{translate}
\end{eqnarray}
resulting in a periodic crystal with two points per fundamental cell.  By construction, 
the number of nearest neighbors to each point in the $d$-dimensional diamond crystal 
is $d+1$, corresponding to one neighbor for each vertex of the regular simplex.  One can verify 
by translation of the fundamental cell that all points of the $d$-dimensional diamond crystal are
topologically equivalent.  Note that the two-dimensional diamond crystal is the 
usual honeycomb lattice, in which each point is at the vertex of a regular hexagon.  

We mention that our construction of the diamond crystal is distinct for all $d\neq 3$ from the $D_d^+$
structure mentioned by Conway and Sloane \cite{CoSl99}.  The $D_d$ lattice is obtained by placing 
points using a ``checkerboard'' pattern in $\mathbb{R}^d$ \cite{CoSl99}:
\begin{eqnarray}
D_d &= \left\{(x_1, \ldots, x_d) \in \mathbb{Z}^d : \sum_{j=1}^d x_j = 2m \mbox{ for some } m \in \mathbb{Z}
\right\}.
\end{eqnarray}
The structure $D_d^+$ is then obtained by including the translate vector 
$\boldsymbol\nu = (1/2, 1/2, \ldots, 1/2)$
in the fundamental cell. Although in three dimensions the $D_d^+$ structure does provide an 
equivalent construction of the diamond crystal, the relationship to our
structure does not hold for any other dimension.  
Indeed, $D_d^+$ is a Bravais lattice for all even dimensions, which is not true for our 
construction of the $d$-dimensional diamond crystal.  For example, in two dimensions, $D_2^+$ is 
equivalent to a rectangular lattice with generator matrix
\begin{eqnarray}
M_{D_2^+} = \left(\begin{array}{cc}
a/2 & 0 \\
0 & a
\end{array}\right),
\end{eqnarray}
where $a$ determines the fundamental cell size.  Each point in this structure possesses two 
nearest neighbors and is therefore distinct from the honeycomb crystal, in which the coordination number 
of each particle is three.  

\subsection{A $d$-dimensional kagom\'e crystal}

\begin{figure}[!t]
\centering
\includegraphics[width=0.325\textwidth]{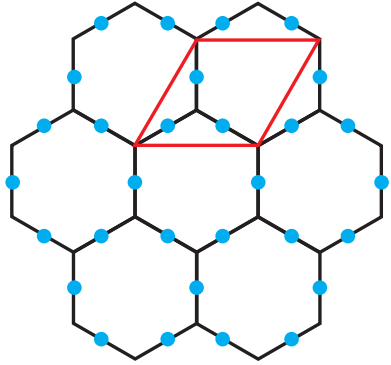}
\caption{Portion of the honeycomb crystal (two-dimensional diamond) with the 
$A_2$ fundamental cell (rhombus).  The points of the honeycomb
crystal are the vertices of the regular hexagons.  The kagom\'e crystal (circular points) 
is then constructed from the
midpoints of the bonds between nearest-neighbors in the honeycomb crystal.  The kagom\'e crystal
consists of vertex-sharing simplices, the centroids of which recover the honeycomb crystal.}\label{Kag2D}
\end{figure}

The two-dimensional kagom\'e crystal is obtained by placing points at the midpoints of 
each nearest-neighbor bond in the honeycomb crystal, resulting in a non-Bravais lattice
with three particles per fundamental cell.  Similarly, the three-dimensional kagom\'e crystal,
also known as the pyrochlore crystal \cite{KaHuUeSc03},
can be constructed by placing points at the midpoints of 
each nearest-neighbor bond in the three-dimensional diamond crystal.  We therefore generalize the 
kagom\'e crystal to higher dimensions using the aforementioned construction of the $d$-dimensional
diamond crystal, placing points at the midpoints of each nearest-neighbor bond.  With respect to 
the underlying $A_d$ Bravais lattice structure, these points are located at
\begin{eqnarray}
\mathbf{x}_0 &= \boldsymbol\nu/2\\
\mathbf{x}_j &= \boldsymbol\nu + \boldsymbol\eta_j/2 \qquad (j = 1, \ldots, d),
\end{eqnarray}
where 
\begin{eqnarray}
\boldsymbol\eta_j &= \mathbf{e}_j-\boldsymbol\nu
\end{eqnarray}
denotes a ``bond vector'' of the $d$-dimensional diamond crystal.  By translating the 
fundamental cell such that the origin is at $\mathbf{x}_0$, we can also represent
the $d$-dimensional kagom\'e crystal as $A_d \oplus \{\mathbf{v}_j\}$, 
where 
\begin{eqnarray}
\mathbf{v}_j &= \mathbf{e}_j /2 \qquad (j = 1, \ldots d).
\end{eqnarray}
The $d$-dimensional kagom\'e crystal therefore has $d+1$ points per fundamental cell, 
growing linearly with dimension.
Each point of the $d$-dimensional kagom\'e crystal is at the vertex of a regular simplex 
obtained by connecting all nearest-neighbors in the structure, implying that each point 
possesses $2d$ nearest neighbors in $d$ Euclidean dimensions \cite{OKe91}.
We illustrate our constructions of the two-dimensional kagom\'e and 
diamond (honeycomb) crystals in Figure \ref{Kag2D}.

\subsection{Other high-dimensional kagom\'e crystals}

Our simple construction of the $d$-dimensional kagom\'e crystal suggests that there exists 
a large family of ``kagom\'e-like'' crystals obtained by including the midpoints of the basis vectors
for a Bravais lattice within the fundamental cell.  A simple example is to include  
basis-vector midpoints into the $d$-dimensional integer lattice $\mathbb{Z}^d$.  A more interesting example is 
obtained in $\mathbb{R}^4$ 
by including the midpoints of the basis vectors for the $D_4$ lattice, the densest
known packing in four dimensions, into the fundamental cell. The packing density of the 
resulting structure is $\phi^{\prime} = 5\pi^2/256 \approx 0.1928$, which should be compared to the 
density of the four-dimensional kagom\'e crystal $\phi = \sqrt{5}\pi^2/128 \approx 0.1724$.  
Note, however, that this kagom\'e-like structure does not possess the same relationship to the
$d$-dimensional diamond crystal as our construction above, and we therefore focus the remainder of the 
discussion on the more natural generalization of the kagom\'e crystal in terms of vertex-sharing simplices
in Euclidean space $\mathbb{R}^d$.

\section{Structural properties of the high-dimensional kagom\'e and diamond crystals}

\subsection{Packing densities and coordination numbers}

The \emph{packing density} associated with a periodic point pattern is the maximal fraction of space
that can be occupied by decorating each of the points with a sphere of radius $R_P$, where 
$R_P$ is the \emph{packing radius}, defined as the maximal value of $R$ for which 
$E_V(R)$ exactly obtains its one-point lower bound at unit number density $\rho$:
\begin{eqnarray}
R_P &\equiv \underset{\rho = 1}{\mbox{sup}}\left\{R : E_V(R) = 1-v(R)\right\}\label{RP}.
\end{eqnarray}
Note that this definition is consistent with our discussion of the packing radius in Section 2.3, involving
decorating each of the points in a point pattern with a sphere of maximal radius $R_P$ such that none of the 
resulting spheres overlap.  However, the definition (\ref{RP}) helps to elucidate the remarkable
connections among the packing, covering, quantizer, and number variance problems. 
For lattices, this formulation is equivalent to identifying the minimal lattice vector at unit number density,
which can be obtained from the in-radius of the Voronoi cell for a given lattice point \cite{CoSl99}.
Note that we have the following weak upper bound on the packing radius for any
Euclidean dimension $d$:
\begin{eqnarray}
R_P &\leq \Gamma^{1/d}(1+d/2)/\sqrt{\pi} \leq R_C,
\end{eqnarray}
where $R_C$ is the covering radius.  Substantially improved upper bounds \cite{CoEl03} and 
conjectural lower bounds \cite{ToSt06} have been provided for the packing radii of the 
densest sphere packings in any Euclidean dimension $d$.

To calculate the packing density of the $d$-dimensional kagom\'e crystal, we first consider the 
$A_d$ Bravais lattice at unit number density, which has a known packing density \cite{CoSl99}
\begin{eqnarray}
\phi_{A_d} &= v\left(R_P^{(A_d)}\right) =  \frac{\pi^{d/2}}{2^{d/2}\Gamma(1+d/2)\sqrt{d+1}},
\end{eqnarray}
where $R_P^{(A_d)}$ is the corresponding packing radius:
\begin{eqnarray}
2R_P^{(A_d)} &= \left(\frac{2^{d/2}}{\sqrt{d+1}}\right)^{1/d}.
\end{eqnarray}
By construction, the $d$-dimensional kagom\'e crystal has the same fundamental cell 
with $d+1$ particles, and the associated packing radius is
\begin{eqnarray}
R_P^{(\mbox{\scriptsize Kag}_d)} &= \left(2^{d/2}\sqrt{d+1} \right)^{1/d}/4 = (d+1)^{1/d} R_P^{(A_d)}/2.
\end{eqnarray}
Therefore,
\begin{eqnarray}
\phi_{\mbox{\scriptsize Kag}_d} &= v\left(R_P^{(\mbox{\scriptsize Kag}_d)}\right) = (d+1) \phi_{A_d}/2^d\\
&= \frac{\pi^{d/2} \sqrt{d+1}}{2^{3d/2} \Gamma(1+d/2)}\label{kagphi}.
\end{eqnarray}

The packing density of the $d$-dimensional diamond crystal can be calculated similarly.  
In particular, the packing radius of the diamond crystal in $d$ Euclidean dimensions is
\begin{eqnarray}
R_P^{(\mbox{\scriptsize Dia}_d)} &= 
\vert\vert\boldsymbol\nu\vert\vert/2 = 2^{1/d} \sqrt{\frac{d}{2(d+1)}} R_P^{(A_d)},
\end{eqnarray}
where $\boldsymbol\nu$ is the translate vector \eref{translate} corresponding to the centroid of the 
regular simplex formed by the basis vectors for the $A_d$ Bravais lattice.  The norm of this 
translate vector can be evaluated by induction and the recursion relation
\begin{eqnarray}
K_d &= \left\vert\left\vert \sum_{j=1}^d \mathbf{e}_j\right\vert\right\vert = \sqrt{K_{d-1}^2 + d}.
\end{eqnarray}
It follows that the packing density of the $d$-dimensional diamond crystal is
\begin{eqnarray}
\phi_{\mbox{\scriptsize Dia}_d} &= 2 \left(\frac{d}{2(d+1)}\right)^{d/2} \phi_{A_d}\\
&= \frac{(\pi d)^{d/2}}{2^{d-1} (d+1)^{(d+1)/2} \Gamma(1+d/2)}.
\end{eqnarray}

\begin{figure}[!t]
\centering
\includegraphics[width=0.5\textwidth]{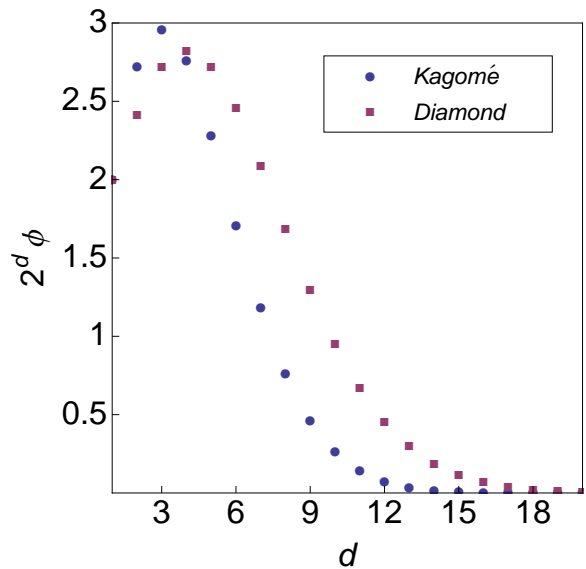}
\caption{Scaled packing densities $2^d \phi$ of 
the $d$-dimensional kagom\'e and diamond crystals.}\label{KDphi}
\end{figure}
Figure \ref{KDphi} compares the packing densities of the $d$-dimensional kagom\'e and 
diamond crystals for increasing dimension $d$.  It is interesting to note that for $d \leq 3$, 
the kagom\'e crystal is a denser packing than the diamond crystal; however, this trend
reverses for all $d \geq 4$.  We will argue in the following sections that this behavior is related 
to the distribution of the void space external to the particles in the lattices.  Specifically, the 
$d$-dimensional kagom\'e structure has increasingly large holes within the fundamental cell,
skewing the void exclusion probability function $E_V(R)$ to higher values of $R$.  This 
behavior implies that the kagom\'e crystal is effectively ``filamentary'' in asymptotically high dimensions,
in the sense that the net of bonds between nearest-neighbors consists of strands that
branch at each point of the crystal but are separated by increasingly large 
holes within the fundamental cell.
While this argument should also hold for the $d$-dimensional diamond crystal, the placement of 
a particle at the centroid of the regular simplex formed by the basis vectors for the fundamental 
cell apparently prevents the lattice holes from growing more rapidly than in the kagom\'e structure.  

We have also determined the coordination numbers for both the $d$-dimensional diamond and 
kagom\'e crystals up to at least the first one hundred coordination shells.  Such calculations 
are helpful in the evaluation of lattice sums for these structures and provide insight into the 
coordination structure of the crystals \cite{ToSt03, OKe91}.
Table \ref{coordtable}  provides abridged results up to 
$d = 5$.
Note that although both structures possess nearest-neighbor coordination numbers growing
linearly with dimension, the kagom\'e crystal in high Euclidean dimensions has a much larger
number of nearest-neighbors than the diamond crystal.  This observation implies that the 
kagom\'e crystal is a much more ``branched'' structure (in
the sense defined above) than the diamond crystal with highly-coordinated
particles separated over increasingly large length scales by holes in the fundamental cell.  
This observation also has implications for the number variance of the kagom\'e structure and 
the decorrelation principle for periodic point patterns, which we discuss in further detail 
in subsequent sections.  
\begin{table}[!t]
\caption{Coordination numbers for the $d$-dimensional diamond (Dia$_d$) and kagom\'e (Kag$_d$) 
crystals.  The square-coordination distance $r_k^2$ is given in parentheses, followed
by the number of neighbor particles $Z_k$ at the distance $r_k$.  The nearest-neighbor
distance determines the length scale for each structure.}\label{coordtable}
\begin{tabular}{c | c c | c c | c c | c c}
\hline\hline
Shell number & Dia$_2$ & Kag$_2$ & Dia$_3$ & Kag$_3$ & Dia$_4$ & Kag$_4$ & Dia$_5$ & Kag$_5$\\
\hline
1 & (1) 3 & (1) 4 & (3) 4 & (4) 6 & (2) 5 & (1) 8 & (5) 6 & (1) 10\\
2 & (3) 6 & (3) 4 & (8) 12 & (12) 12 & (5) 20 & (3) 24 & (12) 30 & (3) 40\\
3 & (4) 3 & (4) 6 & (11) 12 & (16) 12 & (7) 30 & (4) 20 & (17) 60 & (4) 30\\
4 & (7) 6 & (7) 8 & (16) 6 & (20) 12 & (10) 30 & (5) 48 & (24) 90 & (5) 120\\
5 & (9) 6 & (9) 4 & (19) 12 & (28) 24 & (12) 30 & (7) 72 & (29) 90 & (7) 200\\
6 & (12) 6 & (12) 6 & (24) 24 & (32) 6 & (15) 60 & (8) 30 & (36) 140 & (8) 90\\
7 & (13) 6 & (13) 8 & (27) 16 & (36) 18 & (17) 80 & (9) 56 & (41) 240 & (9) 190\\
8 & (16) 3 & (16) 6 & (32) 12 & (44) 12 & (20) 60 & (11) 96 & (48) 270 & (11) 360\\
9 & (19) 6 & (19) 8 & (35) 24 & (48) 24 & (22) 60 & (12) 60 & (53) 210 & (12) 140\\
10 & (21) 12 & (21) 8 & (40) 24 & (52) 36 & (25) 120 & (13) 144 & (60) 360 & (13) 520\\
\hline\hline
\end{tabular}
\end{table}  

\subsection{Void exclusion probabilities, covering radii, and quantizer errors}

As previously mentioned, the covering radius $R_C$ and scaled dimensionless quantizer error 
$\mathcal{G}$ can be determined from knowledge of the void exclusion probability function $E_V(R)$
of a point pattern, which contains information about the distribution of the void space external to the 
particles.  This connection to $E_V$ was first explicitly mentioned recently by Torquato \cite{To10},
and we have been unable to find studies of this function for any periodic crystal in the literature.    
Here, we determine 
$E_V$ for the $d$-dimensional kagom\'e and diamond crystals and use our results to 
provide estimates for the covering radii and quantizer errors for these systems.  

Our calculations involve Monte Carlo sampling of the void space within the fundamental cell for 
the underlying $A_d$ Bravais lattice.  Periodicity of the point pattern implies that $E_V$ must 
have compact support, and it is therefore sufficient only to sample within a single fundamental 
cell, subject to periodic boundary conditions, to obtain the full distribution $E_V$.  Noting that 
any point $\mathbf{r}$ within the fundamental cell can be expressed as an appropriate linear
combination of the Bravais lattice basis vectors:
\begin{eqnarray}
\mathbf{r} &= M_{A_d} \mathbf{x},\label{generator} 
\end{eqnarray}
where $\mathbf{x} = (x_j)_{j=1}^d$ 
with $0 \leq x_j \leq 1$ for all $j$, we can efficiently sample the void space
by placing points randomly and uniformly in the $d$-dimensional unit cube and then mapping those points
to the fundamental cell with the generator matrix $M_{A_d}$ as in \eref{generator}.  The void 
exclusion probability function is then obtained by measuring nearest-neighbor distances between 
the sampling points and the particles of the crystal.  Note that this calculation of the 
void exclusion probability function is more efficient than direct calculation of the 
Voronoi tessellation for the crystals in high dimensions, thereby providing a facile means of 
obtaining estimates for $R_C$ and $\mathcal{G}$.

\begin{figure}[!t]
\centering
$\begin{array}{c c}
\underset{\mbox{\Large(a)}}{\includegraphics[height=2.5in]{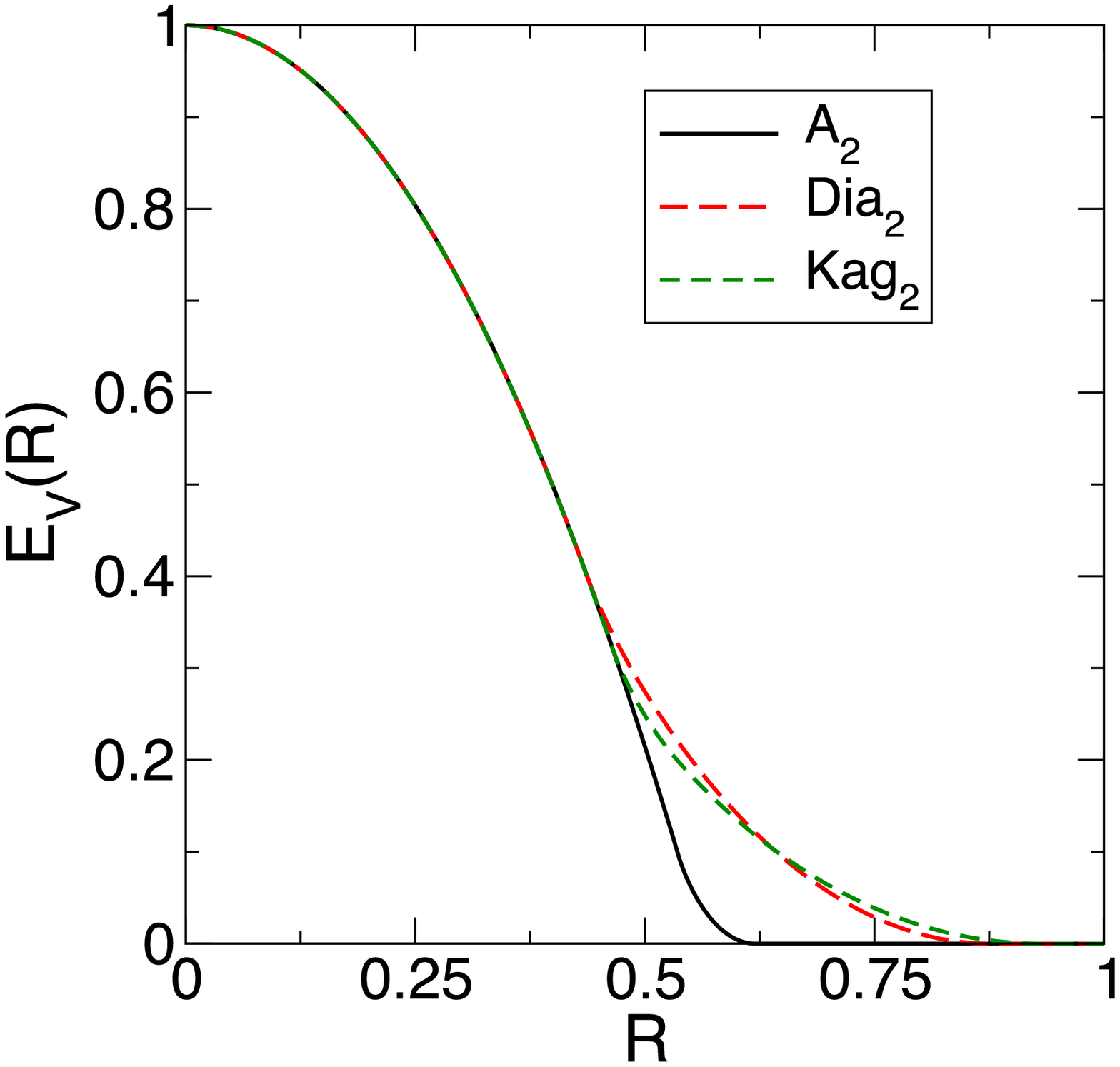}} & 
\underset{\mbox{\Large(b)}}{\includegraphics[height=2.5in]{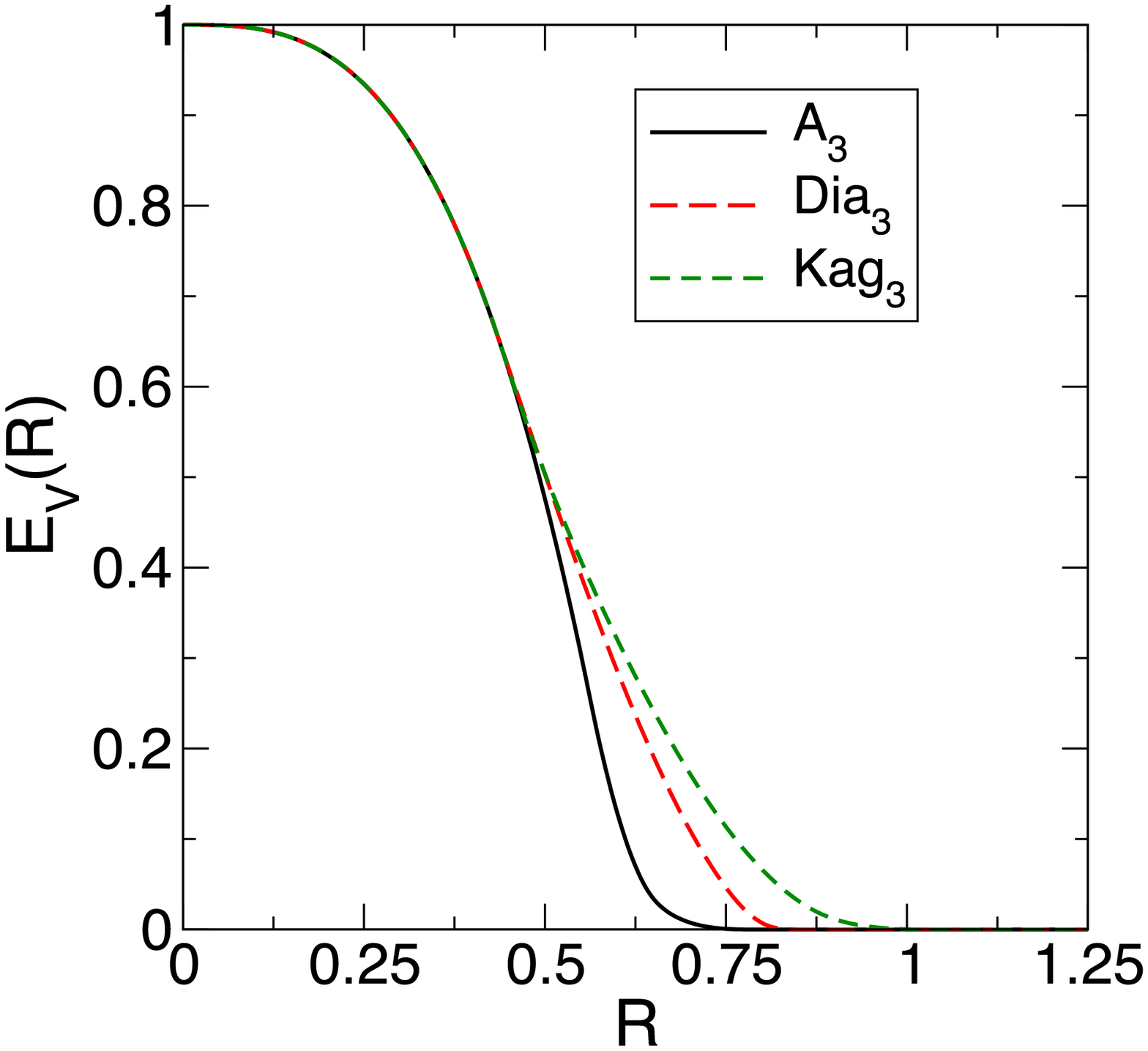}}\\
\underset{\mbox{\Large(c)}}{\includegraphics[height=2.5in]{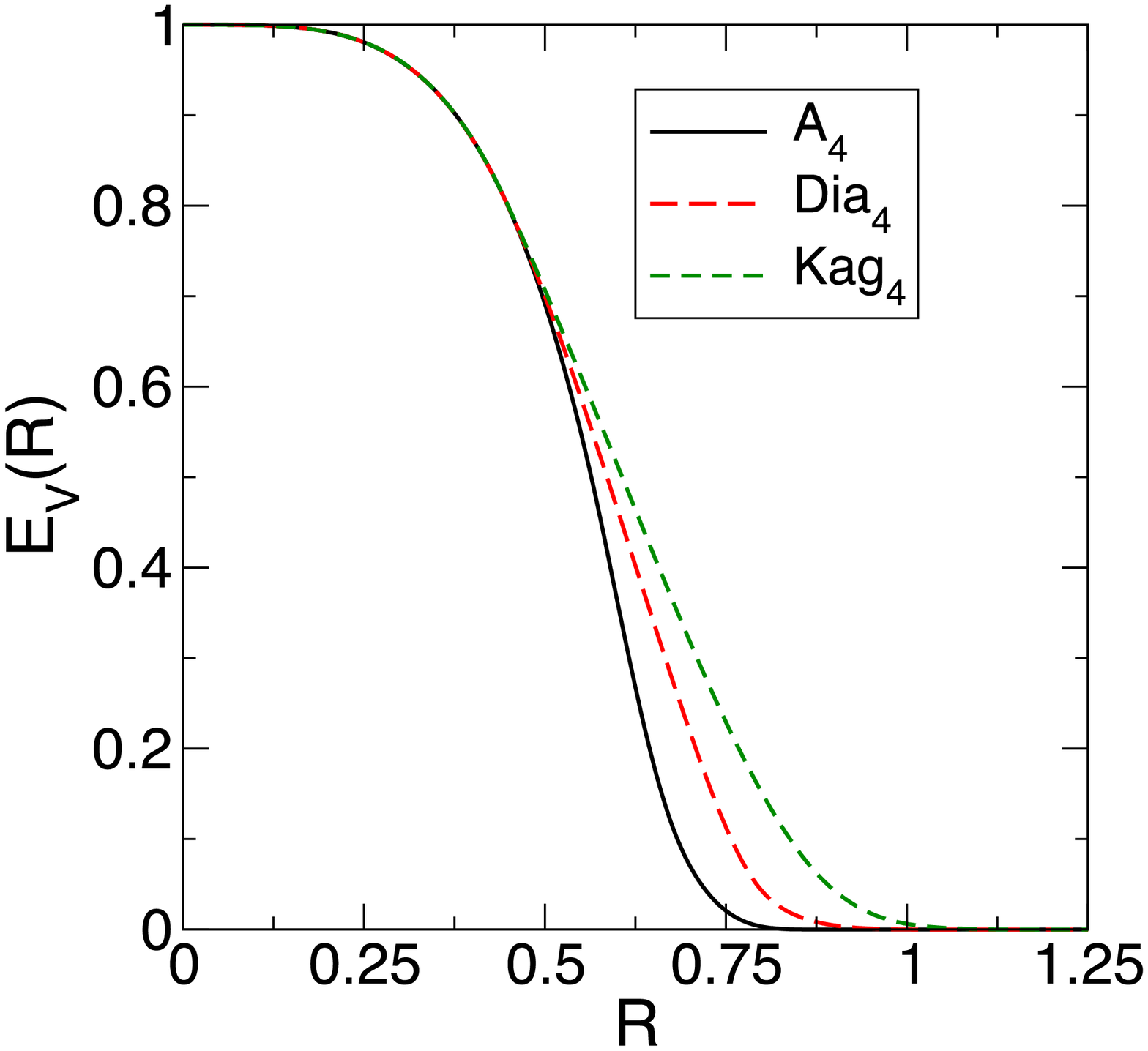}} & 
\underset{\mbox{\Large(d)}}{\includegraphics[height=2.5in]{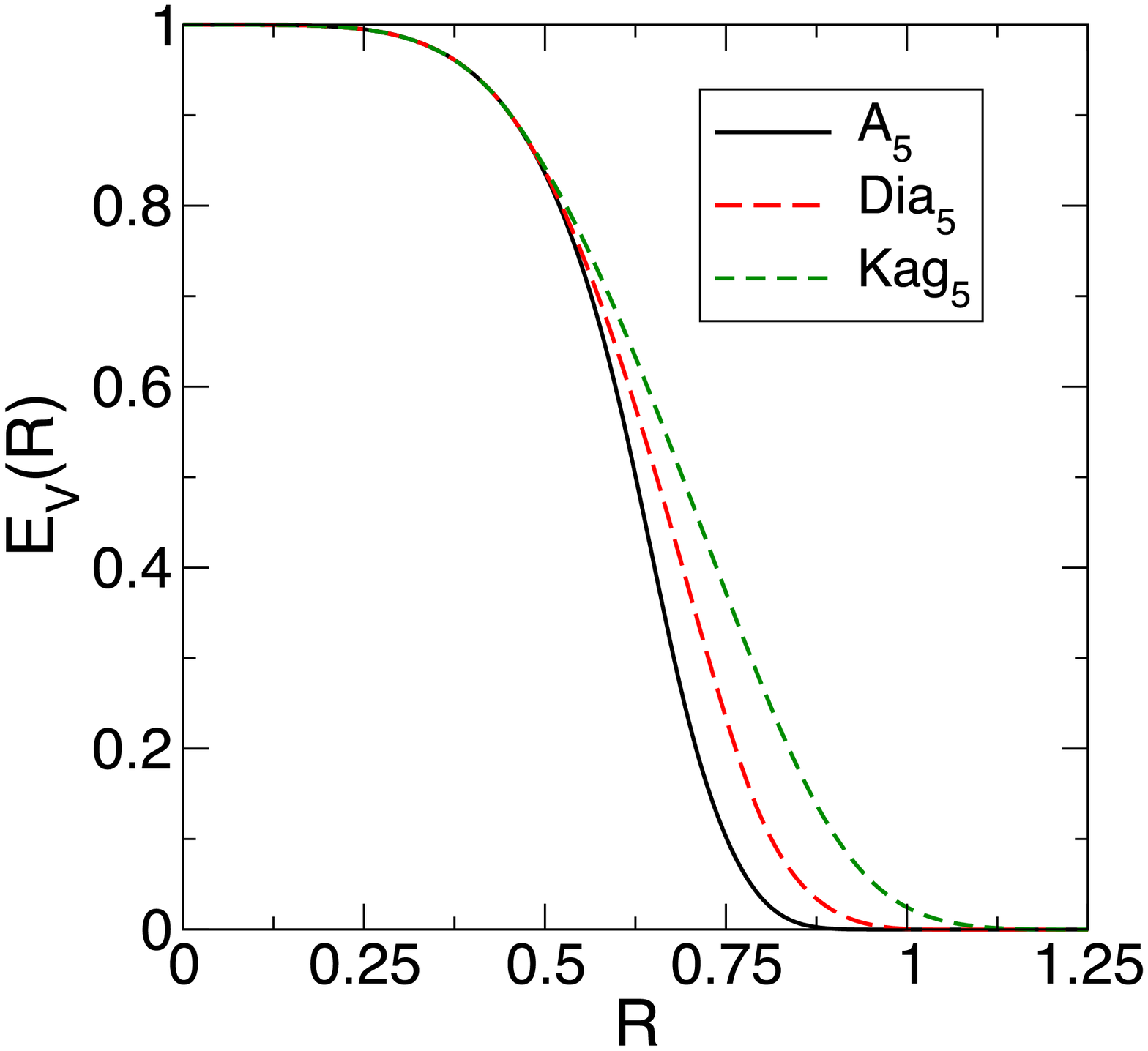}}
\end{array}$
\caption{Void exclusion probability functions $E_V(R)$ for the $A_d$, $d$-dimensional
diamond (Dia$_d$), and $d$-dimensional kagom\'e (Kag$_d$) crystals at unit number
density:  (a) $d = 2$; (b) $d = 3$; (c) $d = 4$; (d) $d = 5$.  }\label{Evd}
\end{figure}
\begin{table}[!t]
\caption{Estimates of the covering radius $R_C$ and quantizer error $\mathcal{G}$
for the $A_d$, $d$-dimensional diamond Dia$_d$, and $d$-dimensional
kagom\'e Kag$_d$ lattices.  Errors for the calculations are $\pm 0.0004$
for the covering radii and $\pm 0.00004$ for the quantizer errors, as 
estimated by comparison with exact results for the $A_d$ lattice in two and three 
dimensions.  The covering radii for the $A_2$, $A_3$, Dia$_2$, and Kag$_2$
lattices are known exactly \cite{CoSl99, To10}, and these exact results are reported here; 
also reported are the exact values for the quantizer errors of the $A_2$ and $A_3$ 
lattices \cite{To10}.}\label{tablecovquant}
\begin{indented}
\item[]\begin{tabular}{c | c | c | c | c | c | c | c | c}
\hline\hline
~ & \multicolumn{2}{c|}{$d = 2$} & \multicolumn{2}{c|}{$d = 3$} & \multicolumn{2}{c|}{$d = 4$} & 
\multicolumn{2}{c}{$d = 5$}\\
\cline{2-9}
~ & $R_C$ & $\mathcal{G}$ & $R_C$ & $\mathcal{G}$ & $R_C$ & $\mathcal{G}$ & 
$R_C$ & $\mathcal{G}$\\
\hline
$A_d$ & 0.6204 & 0.08018 & 0.7937 & 0.07875 & 0.8816 & 0.07780 & 0.9984 & 0.07769\\
\cline{2-9}
Dia$_d$ & 0.8774 & 0.09627 & 0.8640 & 0.09112 & 1.0472 & 0.08825 & 1.0776 & 0.08649\\
\cline{2-9}
Kag$_d$ & 0.9306 & 0.09615 & 1.0384 & 0.09925 & 1.2048 & 0.09973 & 1.2824 & 0.09939\\
\hline\hline
\end{tabular}
\end{indented}
\end{table}
Our results are shown in Figure \ref{Evd}. Table \ref{tablecovquant} summarizes our 
results for the covering radii and quantizer errors of the the diamond and kagom\'e crystals.
The $d$-dimensional kagom\'e crystal possesses relatively large covering radius in 
each dimension, implying that the covering of Euclidean space with the kagom\'e crystal involves
much more than pairwise overlap potentials even in two dimensions.  This behavior 
follows directly from the increasing sizes of holes within the fundamental cell in 
high dimensions.  Since all of the particles in the kagom\'e crystal are relegated to the 
boundary of the fundamental cell, the majority of the space in the fundamental cell is
void space, thereby increasing the value of the covering radius relative to the 
$A_d$ Bravais lattice.  

However, it is interesting to note that the quantizer error for the 
two-dimensional kagom\'e crystal is actually smaller than the associated error for the honeycomb
(two-dimensional diamond) crystal.  Indeed, we recall that kagom\'e crystal generates a denser
sphere packing in two dimensions than the honeycomb crystal, implying by definition that 
$E_V(R) = 1-v(R)$ at unit density for a larger range in $R$.  The void exclusion probability of the 
kagom\'e crystal is therefore relatively ``tight'' compared to the honeycomb crystal in such a way that 
the longer tail does not substantially affect the first moment of the distribution.  The two-dimensional
kagom\'e crystal therefore provides an interesting example of how increasing the complexity of 
a crystal structure can conceivably improve the quantizer error; ``simpler'' structures 
are not always better quantizers, even in low dimensions.  

This behavior changes drastically in higher dimensions, where the quantizer error for the 
$d$-dimensional kagom\'e structure is unusually high relative to the $d$-dimensional diamond 
and $A_d$ structures.  Indeed, the bulk distribution of the void space for the five-dimensional kagom\'e 
crystal is seen to be larger the corresponding curve for a Poisson-distributed point pattern, consisting of 
uncorrelated random points in Euclidean space.  This unusual property implies that the 
quantizer error for the five-dimensional kagom\'e crystal is larger even than Zador's upper bound
\eref{Zador} for the minimal quantizer error.  It is highly counterintuitive that a disordered 
point pattern should be a better quantizer than a periodic crystal with relatively low complexity and 
in relatively low dimensions.  Nevertheless, this observation is consistent with the prevalence of 
large void regions in the high-dimensional kagom\'e crystal and supports our description of this system as
begin effectively ``filamentary'' in high dimensions \cite{FN4}.
This result also suggests the onset of a decorrelation principle 
for the $d$-dimensional kagom\'e crystal, an issue we explore in more detail in Section V.
\begin{figure}[!t]
\centering
\includegraphics[width=0.45\textwidth]{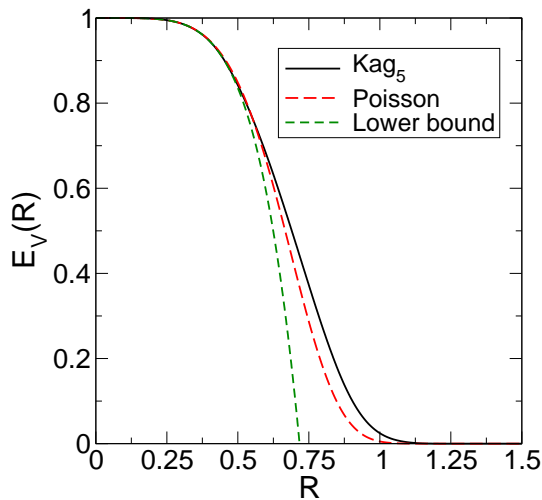}
\caption{Void exclusion probability functions $E_V(R)$ at unit number density for the five-dimensional 
kagom\'e crystal and a disordered, uncorrelated Poisson point process.  Also 
included is the one-point series lower bound $1-v(R)$.}\label{KP}
\end{figure}

\subsection{Number variance coefficients}

As previously mentioned, the asymptotic scaling of the number variance provides a quantitative metric 
for the extent of order within a structure over asymptotically large length scales with respect to 
the mean nearest-neighbor separation between points \cite{ToSt03, ZaTo09}.  Since we are utilizing the 
$d$-dimensional kagom\'e crystal to probe the applicability of the decorrelation principle to 
periodic structures, it is therefore of interest to calculate the surface area coefficient $B$ [c.f.,  \eref{BNV}]
governing surface-area fluctuations in the local number density.  Note that periodicity of the 
fundamental cell implies the presence of full long-range order in both the $d$-dimensional 
kagom\'e and diamond crystals, which is sufficient to induce hyperuniformity.  

Unfortunately, this long-range order also implies that the integral \eref{BNV} diverges; however, 
Torquato and Stillinger have reformulated this expression using a convergence ``trick'' \cite{StDeSa98}
to ensure a properly convergent expression for periodic crystals \cite{ToSt03}.  Specifically, we rewrite
the expression \eref{BNV} for the coefficient $B$ as
\begin{eqnarray}
B &= \lim_{\beta\rightarrow 0^+}\frac{-\rho \kappa(d)}{D} \int \exp(-\beta r^2) r \left[g_2(\mathbf{r})-1\right]
d\mathbf{r},
\end{eqnarray}
where $\kappa(d) = \Gamma(1+d/2)/\{\Gamma[(d+1)/2]\Gamma(1/2)\}$ and $r = \vert\vert\mathbf{r}\vert\vert$.
Expanding this integral implies that
\begin{eqnarray}
B &=\frac{\rho d \pi^{(d-1)/2}}{2 D \beta^{(d+1)/2}} - \frac{\rho \kappa(d)}{D} \int
\exp(-\beta r^2) r g_2(\mathbf{r}) d\mathbf{r} \qquad (\beta\rightarrow 0^+),
\end{eqnarray}
and the remaining integral involving the pair correlation function can be interpreted as the 
average pair sum for the pair interaction $v(r) = \exp(-\beta r^2) r$ over the underlying crystal structure,
which is convergent for all $\beta > 0$.  Writing the average pair sum explicitly, we find
\begin{eqnarray}
\fl B &= \frac{\rho d \pi^{(d-1)/2}}{2D \beta^{(d+1)/2}} - \frac{\kappa(d)}{ND} 
\sideset{}{^\prime}\sum_{j, \ell, \mathbf{p}} \exp\left(-\beta \vert\vert \mathbf{p}+\boldsymbol\nu_j-
\boldsymbol\nu_{\ell}\vert\vert\right)
\vert\vert\mathbf{p}+\boldsymbol\nu_j - \boldsymbol\nu_\ell\vert\vert \qquad (\beta \rightarrow 0^+),
\end{eqnarray}  
where the prime on the summation means that the vector $\mathbf{p}=\mathbf{0}$ is excluded
when $\boldsymbol\nu_j = \boldsymbol\nu_\ell$.  
To remove the dependence of $B$ on the length scale $D$, we 
report the scaled coefficient $\eta^{1/d} B$, where 
$\eta = \rho v(D/2)$, as has previously been done in the literature \cite{ToSt03, ZaTo09}.

\begin{table}[!t]
\caption{Number variance coefficients $\eta^{1/d} B$ for the 
$A_d$, $d$-dimensional kagom\'e Kag$_d$, and $d$-dimensional diamond
Dia$_d$ crystals.  Here, we have taken $\eta$ to be the packing density
of the structure.  The last two entries correspond to $g_2$-invariant processes 
as discussed in the text.  The approximate error for each entry is $\pm 0.00005$
by comparison with previously reported results \cite{ToSt03, ZaTo09}.}\label{Kagnumvar}
\begin{indented}
\item[]\begin{tabular}{c| c c c c c}
\hline\hline
$d$ & $A_d$ & Kag$_d$ & Dia$_d$ & Step-function & Step+delta-function\\
\hline
2 & 0.12709 & 0.14675 & 0.14176 & 0.21221 & 0.15005\\
3 & 0.15569 & 0.20740 & 0.17737 & 0.28125 & 0.19086\\
4 & 0.17734 & 0.27330 & 0.20555 & 0.33953 & 0.22342  \\
5 & 0.19579 & 0.35412 & 0.23144 & 0.39063 & 0.25092\\
\hline
\hline
\end{tabular}
\end{indented}
\end{table}
Table \ref{Kagnumvar} reports our results for the number variance coefficients of the $d$-dimensional
diamond and kagom\'e crystals.  It is helpful to compare these results to similar calculations 
performed for certain so-called \emph{$g_2$-invariant processes} \cite{ToSt02}.  A $g_2$-invariant process
involves constraining a chosen non-negative form for the pair correlation function $g_2$ 
to remain invariant over a nonvanishing density range while keeping all other relevant 
macroscopic variables fixed \cite{ToSt02}.  We consider the following two examples of $g_2$-invariant 
processes:  the so-called ``step-function $g_2$,'' in which the pair correlation function has the 
form
\begin{eqnarray}
g_2(r) &= \Theta(r-D)\label{step}
\end{eqnarray}
for some length scale $D$, and the ``step+delta-function $g_2$'', given by
\begin{eqnarray}
g_2(r) &= \Theta(r-D) + \frac{Z}{\rho s(D)} \delta(r-D),\label{stepdelta}
\end{eqnarray}
where $Z$ can be interpreted as an average contact coordination number \cite{ToSt03}.  Both of these 
processes correspond to \emph{disordered} point patterns that are hyperuniform at the 
critical densities
\begin{eqnarray}
\eta_c &= 1/2^d \qquad (\mbox{step-function})\\
\eta_c &= (d+1)/2^{d+1} \qquad (\mbox{step+delta-function})\label{stepdeta}.
\end{eqnarray}
Strong numerical evidence has been presented to suggest that these pair correlation functions
are indeed realizable as point processes at the critical densities \cite{Uc06}.
Torquato and Stillinger have used $g_2$-invariant processes to define a optimization procedure
that places lower bounds on the maximal sphere-packing density in $d$ Euclidean dimensions 
\cite{ToSt06, ToSt02}.

Torquato and Stillinger have analytically evaluated the number variance coefficients for these
$g_2$-invariant processes:
\begin{eqnarray}
B_{\mbox{\scriptsize step}} &= \frac{d^2 \Gamma(d/2)}{4\Gamma[(d+3)/2]\Gamma(1/2)}\\
2^d \eta_c B_{\mbox{\scriptsize delta+step}} &= 
\frac{d^2 (d+2)\Gamma(d/2)}{16 \Gamma[(d+3)/2]\Gamma(1/2)},
\end{eqnarray}
and these results are included in Table \ref{Kagnumvar}. One notices that for 
all $d\geq 3$, the $d$-dimensional kagom\'e crystal possesses a higher number variance coefficient
than the step+delta-function process, suggesting that there exists a disordered configuration 
of points in high dimensions that is more ordered over asymptotically large length
scales than this periodic structure.  This result is surprising since points in the step+delta-function point
pattern are completely decorrelated from each other beyond the constrained hard-particle diameter.  
Furthermore, the average contact coordination number for this process is \cite{ToSt03}
\begin{eqnarray}
Z &= d/2,
\end{eqnarray}
which is for all dimensions $d$ 
less than the nearest-neighbor coordination number of the $d$-dimensional
kagom\'e crystal, $Z_{\mbox{\scriptsize Kag}_d} = 2d$.  

To understand this behavior, we first note that the packing density \eref{stepdeta} is less than 
the corresponding density \eref{kagphi} for the $d$-dimensional kagom\'e crystal for all $d \leq 4$;
however, for $d \geq 5$, the step+delta-function process possesses a higher packing density
than the kagom\'e crystal. This observation implies that the local ordering between points induced by the 
delta-function contribution to the pair correlation function of the step+delta-function process
is sufficient to regularize the void space in such a way that the packing radius $R_P$
remains relatively high compared the kagom\'e structure.  In particular, the large holes within 
the kagom\'e fundamental cell control the structural properties of the point pattern in high
Euclidean dimensions, and it is these holes that increase the asymptotic number variance coefficient
in such a way that the point pattern can no longer be distinguished from correlated but disordered 
point patterns.  This behavior is in accordance with an effective decorrelation between the 
points of the kagom\'e structure over large length scales and supports the presence of a 
decorrelation principle for this system.  

It is important to note that the increasing nearest-neighbor coordination number of the $d$-dimensional 
kagom\'e crystal implies that correlations between nearest-neighbors are \emph{increasing} 
with increasing dimension.  For this reason, the number variance coefficient governing surface 
area fluctuations is always smaller than the corresponding coefficient for the simple step-function 
process in any dimension; these \emph{constrained} correlations are never removed by the
dimensionality of the system.  However, correlations over several nearest-neighbor distances
apparently diminish in an effective manner, which we make more precise in Section V, 
and it is this type of decorrelation that we claim is responsible for unusually large 
asymptotic local-number-density fluctuations in the kagom\'e structure.  Note also that these
results are consistent with our analysis of the quantizer errors for the $d$-dimensional kagom\'e crystals.

\section{The decorrelation principle for periodic point patterns}

\subsection{Universality of decorrelation in high dimensions}
The decorrelation 
principle \cite{ToSt06} states that 
unconstrained asymptotic $n$-particle correlations vanish in sufficiently 
high dimensions, and all higher-order $(n\geq 3)$ correlation functions 
can be expressed in terms of the pair correlation function within some small error.  
Although originally stated in the context of hard-sphere packings, certain ``soft'' 
many-particle distributions, including points interacting in the Gaussian core model \cite{Za08}
and noninteracting spin-polarized ground-state fermions \cite{To08}, are also known to exhibit this effect,
even in relatively low dimensions $d = 1$-$6$.  
No rigorous proof for this principle has been found to date, but
it is based on strong theoretical arguments and has been shown to be 
remarkably robust in theoretical and numerical studies.  

Does the decorrelation principle apply in some generalized sense to periodic crystals?
It is not trivial to extend the decorrelation principle to periodic crystals,
which inherently possess long-range order owing to the 
regular arrangement of points within a lattice structure. 
This full long-range order induces 
deterministic correlations as manifested by Bragg peaks in the power spectrum.
In particular, we recall from \eref{g2periodic} that the angularly-averaged pair correlation function
consists of consecutive Bragg peaks at each coordination shell; it is convenient to 
express this relation in terms of the packing fraction $\phi$ and associated packing diameter $D$:
\begin{eqnarray}
g_2(r) &= \sum_{k=1}^{+\infty} \frac{Z_k}{2^d d \phi}\left(\frac{D}{r_k}\right)^{d-1} \delta[(r-r_k)/D].\label{pcf}
\end{eqnarray}
The intensity of each peak in the pair correlation function is therefore determined by the 
coordination number $Z_k$ of the $k$-th coordination shell, the packing density $\phi$, 
and the distance $r_k$ to the $k$-th coordination shell.  

\begin{figure}[t]
\centering
\includegraphics[width=0.5\textwidth]{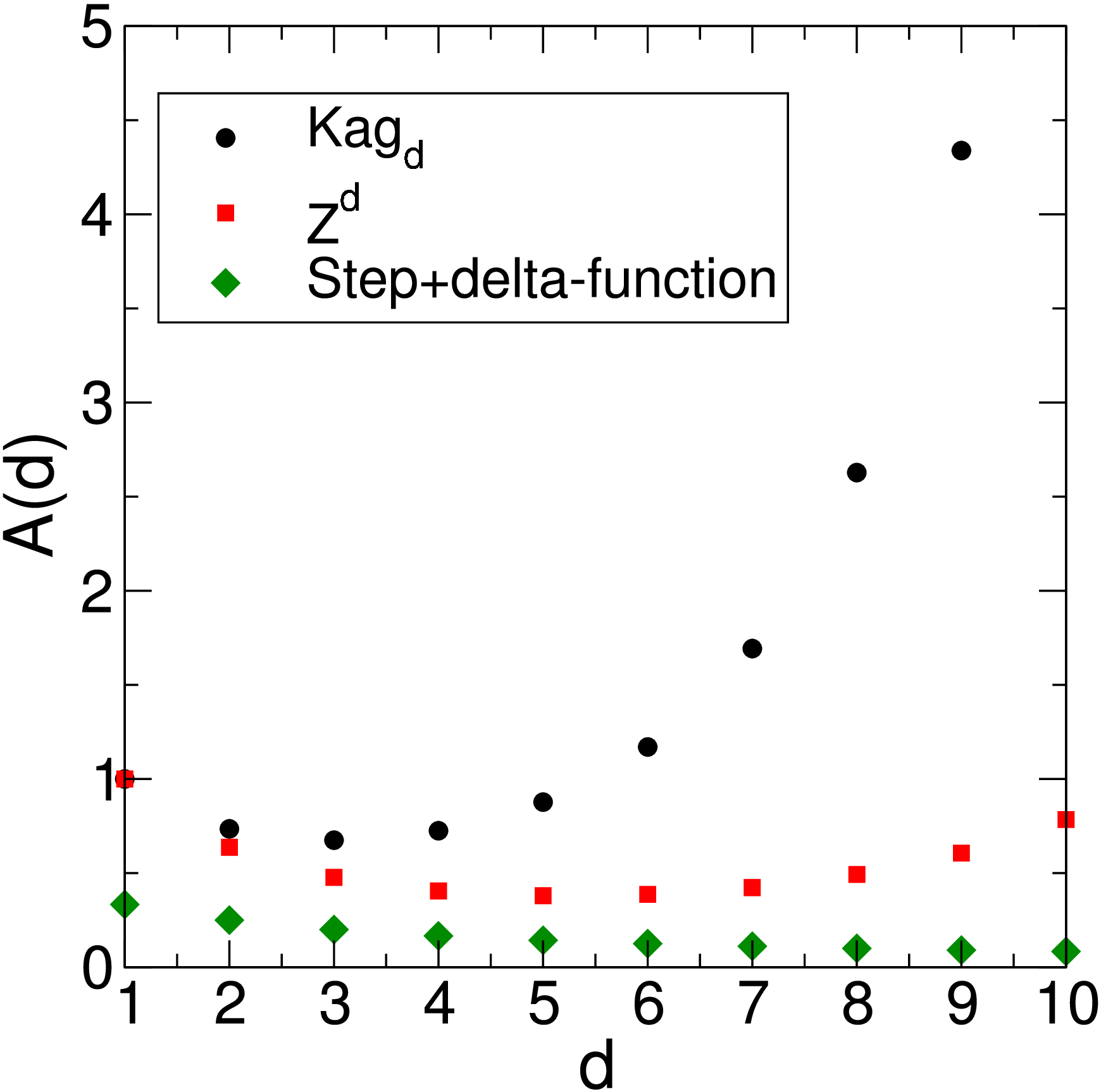}
\caption{Intensity $A(d) = Z_1(d)/(2^d d \phi)$ associated with the first $\delta$-function peaks
of the pair correlation functions for the $d$-dimensional kagom\'e and hypercubic 
$\mathbb{Z}^d$ crystals.  Also shown for comparison is the result for the 
$g_2$-invariant step+delta-function process \eref{stepdelta}.}\label{figAd}
\end{figure}

It is interesting to examine the behavior
of the intensity 
\begin{eqnarray}
A(d) = Z_1(d)/(2^d d \phi)\label{AD}
\end{eqnarray}
associated with the first peak of the pair correlation function 
for the $d$-dimensional kagom\'e and hypercubic $\mathbb{Z}^d$ crystals, shown in Figure \ref{figAd}.
Note that both of these crystals possess a nearest-neighbor contact number $Z_1(d) = 2d$, 
equivalent to the isostatic condition \cite{FNiso}.
After an initial drop in relatively low dimensions, the intensity $A(d)$ increases without bound for 
both of the periodic systems.  Furthermore, the $d$-dimensional kagom\'e crystal possesses a 
first-shell intensity that grows much more rapidly with dimension than even the 
hypercubic lattice $\mathbb{Z}^d$, 
which is a direct consequence of the exponentially diminishing packing density
and the prevalence of large holes in the fundamental cell.  In both cases, nearest-neighbor
correlations asymptotically \emph{increase} with dimension $d$, and it is therefore 
unclear whether a decorrelation principle should hold for periodic crystals.
This behavior should be contrasted 
with corresponding results for the disordered $g_2$-invariant step+delta-function process 
\eref{stepdelta}, where the first-peak intensity $A(d) = 1/(d+2)$ diminishes 
for all dimensions $d$.

Nevertheless, one can consider a disordered point pattern to be a 
realization of a non-Bravais lattice with a large number of particles
randomly distributed in the fundamental cell.  This observation suggests 
that periodic crystals with an $M$-particle basis  
growing with dimension may exhibit the same decorrelation properties as a 
disordered many-particle distribution.  If this notion is true, then 
the effects of decorrelation should then be readily observed 
even in relatively low dimensions as with disordered packings 
\cite{ToSt06, To06B, ToUcSt06, Za08, To08, Sk06}.
It is therefore intriguing to test the 
decorrelation principle for the $d$-dimensional kagom\'e lattice, which, as previously mentioned,
possesses $d+1$ particles per fundamental cell.  

The deterministic long-range order of a periodic crystal implies that the decorrelation principle,
if it applies, cannot be directly observed from the pair correlation function \eref{pcf} itself,
but rather from some smoothed form of $g_2$.
Instead, the pair correlation function of a crystal must be interpreted in the sense of 
\emph{distributions} \cite{Li01}; it gains physical meaning only when integrated 
with an admissible function.  Therefore, the appropriate function to consider is the
``smoothed'' pair correlation function
\begin{eqnarray}
g_2^{(a)}(r) &= \sum_{k=1}^{+\infty} \frac{Z_k D}{2^{d} d \phi a \sqrt{\pi}} \left(\frac{D}{r_k}\right)^{d-1}
\exp\left\{-[(r-r_k)/a]^2\right\},\label{g2smooth}
\end{eqnarray}
corresponding to a convolution of \eref{pcf} with a Gaussian kernel \cite{Li01, Bl93}.
Note that $g_2^{(a)}(r) \rightarrow g_2(r)$ for $r \in [0, +\infty)$ 
(in the sense of distributions) as $a\rightarrow 0$.  

\begin{figure}[t]
\centering
\includegraphics[width=0.5\textwidth]{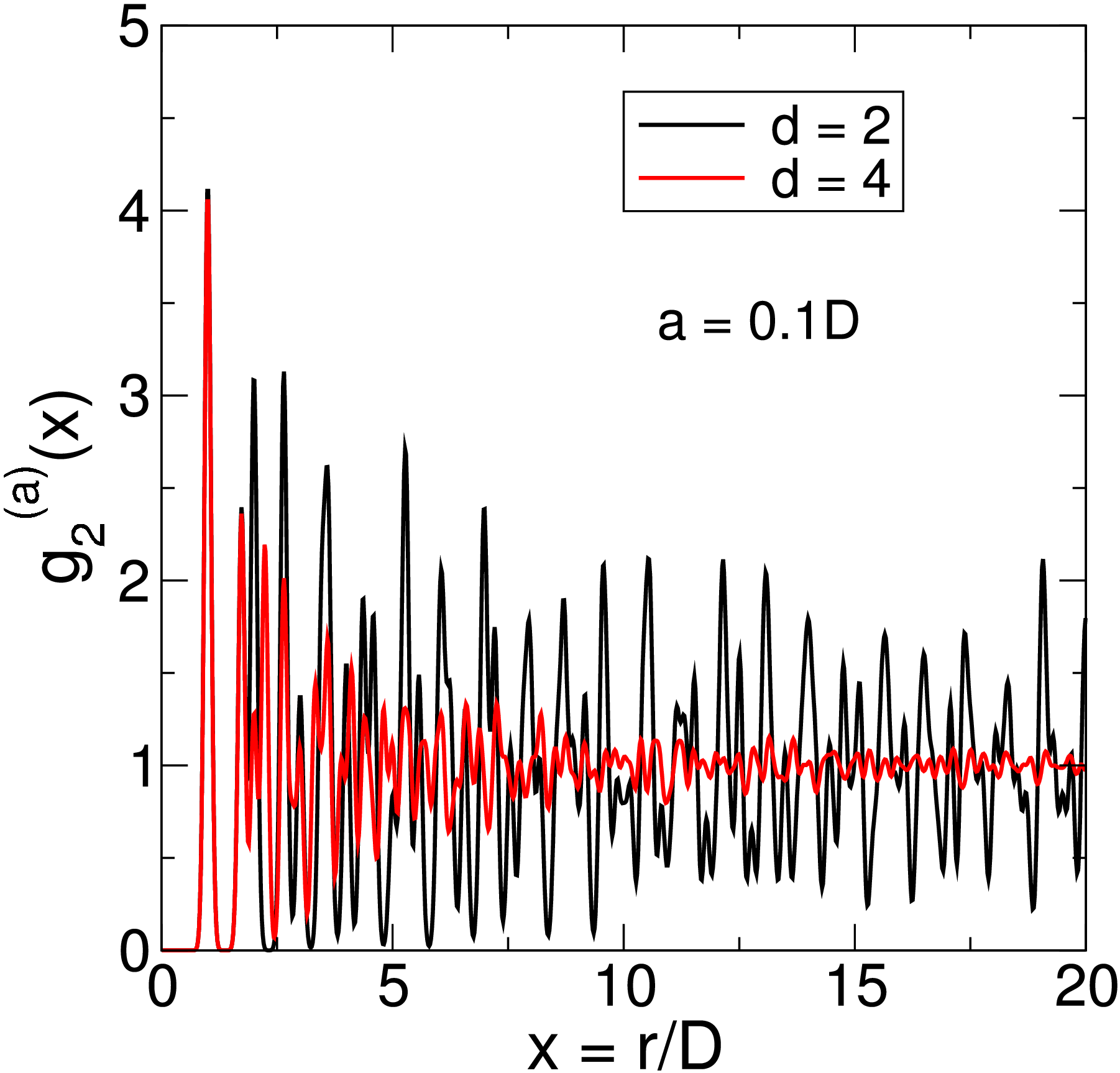}
\caption{Smoothed pair correlation function for the kagom\'e crystal in dimensions
$d = 2$ and $d = 4$.  The smoothing parameter $a = 0.1D$ [c.f. \eref{g2smooth}].}\label{Kagg2ref}
\end{figure}
Figure \ref{Kagg2ref} compares the smoothed pair correlation functions for the $d = 2$ and $d = 4$
kagom\'e lattices.  Remarkably, asymptotic pair correlations are observed to diminish 
even in the relatively low dimensions shown (as with 
disordered point patterns \cite{ToSt06, Za08, To08, Sk06, To06B, ToUcSt06}),
implying that the pair correlation function
approaches its asymptotic value of unity in sufficiently high dimensions.    Importantly, 
this effect at large pair separations is observed for any nonzero choice of the smoothing parameter $a$ with only qualitative differences in the pair correlation function, corresponding to localization 
of the $\delta$-function peaks.  Our results therefore suggest that the decorrelation principle 
applies to the $d$-dimensional kagom\'e crystal in the sense that \emph{any} delocalization 
of the local density field is sufficient to cause asymptotic pair correlations to diminish 
with respect to increasing dimension.  Note that these observations are consistent with 
our calculations for the asymptotic number variance coefficient for the kagom\'e crystal, 
which is higher than the corresponding result for the disordered step+delta-function process 
even in low dimensions.  

\begin{figure}[t]
\centering
\includegraphics[width=0.5\textwidth]{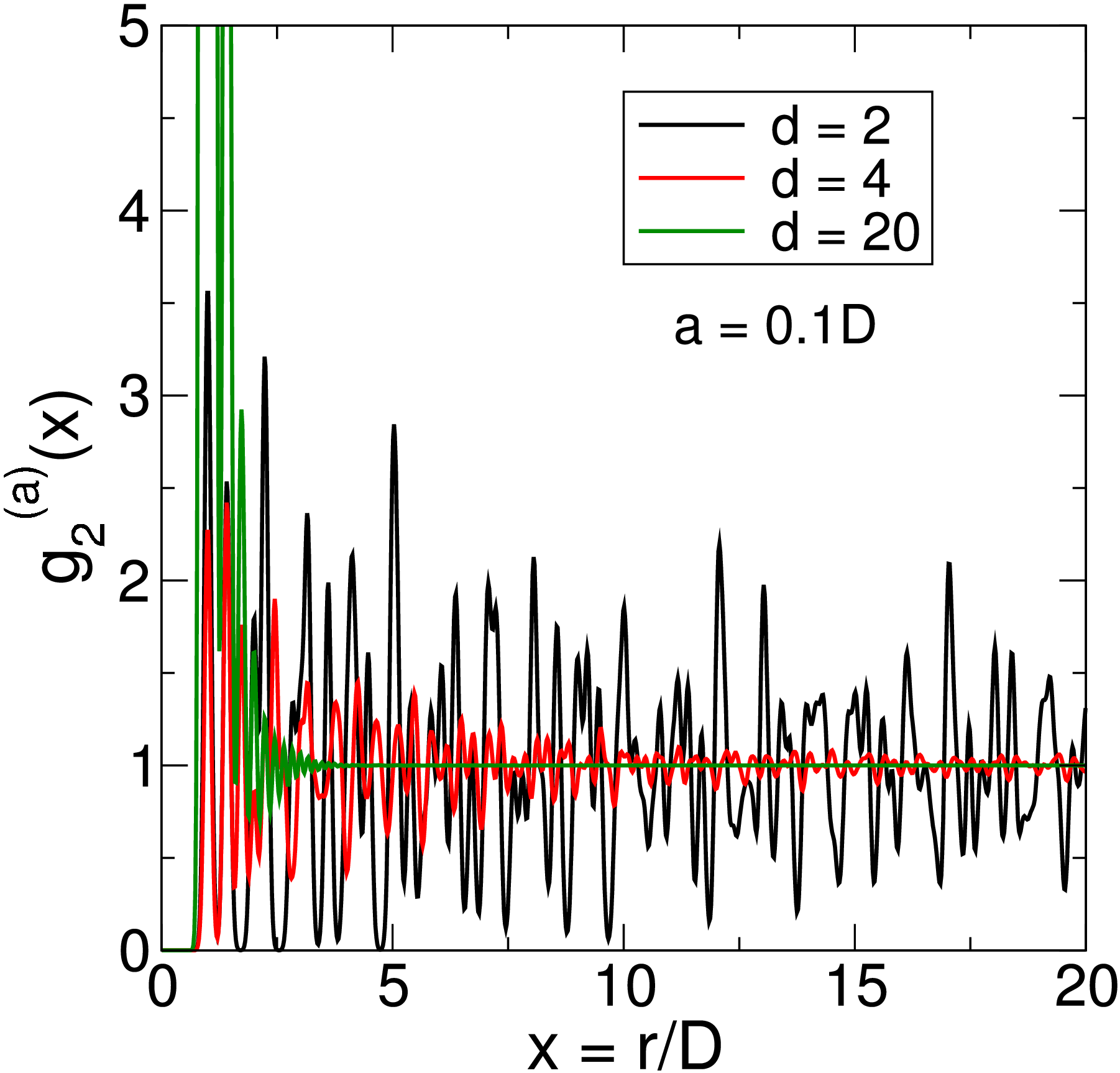}
\caption{Smoothed pair correlation function for the hypercubic Bravais lattice $\mathbb{Z}^d$ 
in dimensions
$d = 2$, $d = 4$, and $d = 20$.  The smoothing parameter $a = 0.1D$.}\label{Zg2}
\end{figure}
Additional calculations suggest that this approach to the decorrelation principle for periodic structures 
is applicable even to crystals with $M$-particle bases that do not increase with dimension.  Figure 
\ref{Zg2} provides calculations of the smoothed pair correlation function for the hypercubic Bravais lattice
$\mathbb{Z}^d$.  Like the kagom\'e crystal, decorrelation is readily apparent even in low dimensions, 
and upon reaching $d = 20$ the system is essentially completely decorrelated beyond 
a few nearest-neighbor distances.   
Since any point pattern, disordered or not, can be modeled as a periodic point pattern, potentially
with a large number of points per fundamental cell, these observations support the remarkable statement
that the decorrelation principle is a \emph{universal} feature of high-dimensional 
point patterns, including those distributions associated with sphere packings.  
In particular, the principle should apply not only to disordered point patterns as originally 
discussed by Torquato and Stillinger \cite{ToSt06} but also to periodic crystals.

The smoothing operation that we have introduced for the
pair correlation function allows us to observe the effects of decorrelation in periodic 
crystals in relatively low dimensions.  In asymptotically high dimensions, 
the widths of the Gaussians can be made 
arbitrarily small since consecutive coordination shells are tightly 
clustered.  
Decorrelation is therefore a fundamental feature of the pair correlation function itself
of a high-dimensional periodic point pattern, whether it is a simple Bravais lattice 
or a crystal with many points per fundamental cell.  This principle supports the claim that
higher-order correlation functions do not provide additional information beyond 
that contained in $g_2$, meaning that the pair correlations alone 
completely determine the packing 
in high dimensions.   


\subsection{Implications for the maximal density of sphere packings}

The onset of decorrelation in high dimensions for periodic crystals has important implications for 
optimal lower bounds on the maximal sphere-packing density.  Minkowski provided 
a \emph{nonconstructive} proof that the asymptotic behavior of the maximal density of 
sphere packings is bounded from below by \cite{ToSt06, Mi05}
\begin{eqnarray}
\phi_{\mbox{\scriptsize max}} &\gtrsim \frac{1}{2^d} \qquad (d\rightarrow +\infty).
\end{eqnarray}
This scaling is quite distinct from the Kabatiansky-Levenshtein upper bound on the
maximal sphere packing density \cite{KaLe78}
\begin{eqnarray}
\phi_{\mbox{\scriptsize max}} &\leq \frac{1}{2^{0.5990d}} \qquad (d\rightarrow +\infty).\label{KL}
\end{eqnarray}
Utilizing the decorrelation principle for \emph{disordered} sphere packings, Torquato and 
Stillinger derived a conjectural lower bound on the maximal sphere-packing density that provides
putative exponential improvement over Minkowski's bound \cite{ToSt06}:
\begin{eqnarray}
\phi_{\mbox{\scriptsize max}} &\gtrsim \frac{d^{1/6}}{2^{0.77865\ldots d}} \qquad (d\rightarrow +\infty).\label{TS}
\end{eqnarray}
This bound was derived using the aforementioned $g_2$-invariant optimization procedure
for a ``test'' pair correlation function that in the high-dimensional limit becomes a step+delta 
function.  It is a conjectural bound because it has yet to be shown that such a pair correlation 
function with packing density \eref{TS} is realizable by a point process, an issue to which we 
will return.  
The gap between the Kabatiansky-Levenshtein upper bound and the Torquato-Stillinger lower bound
remains relatively large in high dimensions, and it is therefore an open problem to determine
which bound provides the ``correct'' asymptotic scaling.

To gain some insight into this issue, we note that
in sufficiently high 
dimensions the distances between subsequent coordination shells become increasingly 
small, implying that the smoothing parameter $a$ used to observe the decorrelation effect 
in the pair correlation function does not need to be chosen very large.  In the asymptotic dimensional
limit, it follows that \emph{any} choice of the smoothing parameter is sufficient to ``collapse'' 
the pair correlation function onto its asymptotic value of unity with the exception of nearest-neighbor
correlations, which are dominant in high dimensions.  We therefore emphasize that 
the smoothing operation we have employed in this work is only a convenient tool 
that allows us to observe the decorrelation principle in even relatively low dimensions.  The decorrelation
principle itself is apparently a \emph{fundamental} and \emph{universal} phenomenon of 
any high-dimensional point pattern, ordered or not, 
manifested in the pair correlation function since higher-order correlation functions 
do not introduce additional information \cite{ToSt06}.  In summary,
decorrelation suggests that the pair correlation functions
of general periodic point patterns tend to the step+delta-function form, which is precisely 
the same asymptotic form as the test function that Torquato and Stillinger used to obtain 
the lower bound \eref{TS} \cite{ToSt06, ScStTo08}.
Appendix A contains an analytical demonstration of the decorrelation 
effect for the previously-mentioned $g_2$-invariant step-function and step+delta-function processes.

However, the asymptotic scaling of the packing density for a sphere packing 
will depend inherently on the \emph{manner} in which the pair correlation function approaches
this asymptotic form.  
Since the dominant correlations in asymptotically high dimensions will be from nearest-neighbors 
in a sphere packing, owing to the well-defined exclusion region in the pair correlation function, 
the decorrelation principle suggests that all sphere packings in high dimensions possess
pair correlation functions of an effective step+delta-function form, which is 
precisely the same asymptotic form as the test function that Torquato and 
Stillinger used to obtain the lower bound \eref{TS} \cite{ToSt06, ScStTo08}.  The intensity $A(d)$ of the 
associated delta-function peak is given by (\ref{AD}).  Whether this intensity 
increases (as with the $d$-dimensional 
kagom\'e and hypercubic $\mathbb{Z}^d$ crystals)
or diminishes [as with the $g_2$-invariant 
step+delta-function process at its critical density \eref{stepdeta}] 
in asymptotically high dimensions therefore depends on the 
relative scalings of $Z(d)$ and $\phi(d)$.  

Using the same linear programming techniques introduced in Ref. [12], 
Scardicchio, Stillinger, and Torquato have numerically explored the 
$Z(d)$-$\phi(d)$ parameter space associated with the step+delta-function process
when hyperuniformity is \emph{not} enforced \emph{a priori} 
[as it is at the critical density \eref{stepdeta}] \cite{ScStTo08}.  Their results provide 
the \emph{same} exponential improvement on Minkowski's lower bound for the 
maximal sphere-packing density as the Torquato-Stillinger lower bound \eref{TS}.  
Additionally, this scaling is robust in the sense that it is recovered for 
test pair correlation functions containing any number of delta-function peaks \cite{ScStTo08}.
This latter observation implies that next-nearest-neighbor correlations, even if they 
persist in high dimensions, do not provide additional exponential improvement 
over the Torquato-Stillinger lower bound \eref{TS} for the maximal sphere-packing density.    

Since the $g_2$-optimization procedure identifies the maximal packing density 
obtainable with the step+delta-function form, which is also 
obtained by high-dimensional periodic sphere packings by the decorrelation principle,
our results support the remarkable possibility that the Torquato-Stillinger
lower bounds may in fact be \emph{optimal} in asymptotically high dimensions.  If confirmed,
this result would imply that the 
Kabatiansky-Levenshtein upper bound \eref{KL} therefore provides a suboptimal
high-dimensional estimate.  
This conclusion is consistent with similar arguments put forth in 
Ref. [51].  If this claim is true, it is interesting to note that the results of 
Scardicchio, Stillinger, and Torquato suggest that the intensity of the nearest-neighbor peak in
the pair correlation function of maximally dense sphere packings in high dimensions 
should \emph{diminish} with increasing $d$ \cite{ScStTo08}, implying that the full pair correlation function 
completely decorrelates to a step-function form.  It follows that certain periodic point patterns such as the 
$d$-dimensional kagom\'e and hypercubic $\mathbb{Z}^d$ crystals \emph{cannot} be 
maximally dense in high dimensions, thereby 
providing direct evidence that the manner in which 
the pair correlation function approaches the asymptotic 
step+delta-function form reflects the high-dimensional asymptotic
scaling of the packing density.  Indeed, the maximally dense sphere packings 
in high dimensions may therefore likely be disordered
(i.e., with a pair correlation function decaying to unity sufficiently 
fast in the infinite-volume limit \cite{ToSt06}) as first suggested by Torquato and 
Stillinger \cite{ToSt06}.

\section{Concluding remarks}

We have provided constructions of the high-dimensional generalizations of 
the kagom\'e and diamond crystals.  The $d$-dimensional diamond crystal is obtained by 
including in the $A_d$ fundamental cell the centroid of the regular simplex formed by the 
lattice basis vectors.  The $d$-dimensional kagom\'e crystal can then be constructed by 
placing points at the midpoints of the ``bonds'' in the diamond crystal.  The kagom\'e crystal 
possesses a nearest-neighbor contact number $Z = 2d$ in $d$ Euclidean dimensions, which is
equivalent to the isostatic condition for jammed sphere packings.  
In two dimensions, the kagom\'e crystal is locally but neither collectively nor strictly jammed \cite{Do04} under periodic boundary conditions; 
however, it can be reinforced to obtain the lowest-density strictly jammed 
subpacking of the triangular lattice.  In three dimensions, 
the pyrochlore crystal has clustered equilateral-triangle vacancies.  In contrast 
to $d = 2$,  the $d$-dimensional 
kagom\'e crystal is therefore not strictly jammed for any $d\geq 3$ \cite{OKe91}.

Using these constructions, we have derived analytically the packing densities of these structures
and have shown that while the kagom\'e crystal generates a denser sphere packing for $d = 2$ 
and $d = 3$, the diamond crystal is denser for all $d \geq 4$, at which point the holes 
in the kagom\'e lattice become substantially large.  These observations are supported by 
numerical calculations for the void exclusion probabilities of the kagom\'e and diamond crystals.
Surprisingly, the bulk of the void-space distribution for the kagom\'e lattice in dimensions $d \geq 5$
is larger than the corresponding result for the disordered, fully-uncorrelated Poisson point pattern.

Our results have implications for the quantizer errors and covering radii of these structures in high
dimensions.  The diamond crystal provides a thinner covering of space than the kagom\'e crystal 
for all $d\geq 2$, even though the kagom\'e crystal is a better quantizer in two dimensions.  
However, the large holes in the fundamental cell for the kagom\'e lattice rapidly increase 
its quantizer error in high dimensions such that it even exceeds Zador's upper bound
and, therefore, also Torquato's improved upper bound \cite{To10}, in as low as $d = 5$.
This observation implies that disordered point patterns can be better quantizers than certain 
periodic structures even in relatively low dimensions, which is consistent with the 
properties of certain disordered point patterns reported in Ref. [1].  

We have also calculated the asymptotic surface-area coefficients for the number variance of the 
kagom\'e and diamond crystals.  Interestingly, the $d$-dimensional kagom\'e lattice possesses a larger
asymptotic number variance coefficient even than the disordered step+delta-function $g_2$-invariant
process for all $d \geq 3$.  Since the number variance coefficient provides a quantitative measure
of structural order over large length scales  \cite{ToSt03, ZaTo09}, this result counterintuitively 
suggests that periodic crystals may possess \emph{less} long-range structural order than 
prototypical ``disordered'' point patterns in sufficiently high dimensions, which is consistent with 
a generalized decorrelation principle for periodic structures.   
By calculating a ``smoothed'' pair correlation function for the $d$-dimensional kagom\'e 
crystal, we have provided direct evidence for the decorrelation principle in periodic 
point patterns.  Indeed,
the decorrelation principle appears to be \emph{universal}, applying also to Bravais lattices
as shown by corresponding calculations for the hypercubic lattice $\mathbb{Z}^d$ in high dimensions.  

Our work has important implications for the maximal sphere-packing density in high Euclidean 
dimensions.  In particular, the suggested universality of the decorrelation principle for both disordered 
and periodic sphere packings suggests that the putative exponential improvement 
obtained by Torquato and Stillinger \cite{ToSt06} on 
Minkowski's lower bound for the maximal packing density is in fact optimal, which is consistent 
with previously-reported results in the literature \cite{ScStTo08}.  The pair correlation functions of
high-dimensional sphere packings apparently possess a general step+delta-function form, and 
optimization of the packing structure through the 
$Z$-$\phi$ parameter space \cite{ScStTo08}, where $Z$ is the 
mean nearest-neighbor contact number and $\phi$ is the packing density, suggests that 
maximally dense packings undergo a complete decorrelation in high Euclidean dimensions.  
In particular, the intensity of the nearest-neighbor peak in the pair correlation function diminishes 
in high dimensions, which should be contrasted with the corresponding behaviors for the 
$d$-dimensional kagom\'e and hybercubic $\mathbb{Z}^d$ crystals.  These latter structures 
therefore cannot be maximally dense in high dimensions, which is in accordance with the 
notion that the densest packings for asymptotically large $d$ are in fact disordered \cite{ToSt06}.
Importantly, this work provides the foundation for a rigorous proof of the 
Torquato-Stillinger lower bound on the maximal-sphere packing density and its optimality
in high dimensions.
Future work is also warranted to explore the implications of the decorrelation 
principle for the covering, quantizer, and number variance problems.

\ack
This work was supported by the Materials Research Science and Engineering Center
(MRSEC) Program of the National Science Foundation under Grant No. 
DMR-0820341.  

\appendix

\section{Analytical demonstration of the decorrelation principle for a simple example}

Here we examine the step-function and step+delta-function $g_2$-invariant processes 
at their critical densities [c.f. \eref{step}-\eref{stepdeta}].  Our goal is to probe the
manifestation of the decorrelation principle in delta-function contributions to the pair correlation function. 
As mentioned in the text, the step+delta-function $g_2$-invariant process is expected to 
provide a good approximation to the effective pair correlation function of periodic crystals 
in asymptotically high Euclidean dimensions.   
Consider the linear average of $g_2$ over some radial length scale $L$:
\begin{eqnarray}
G(L) = \frac{1}{L}\int_0^{L} g_2(r) dr.
\end{eqnarray}
For the step-function process, this linear average is given by
\begin{eqnarray}
G_{\mbox{\scriptsize step}}(L) = \left(1-\frac{D}{L}\right)\Theta(L-D)
\end{eqnarray}
for all dimensions.  By contrast, for the step+delta-function process, we find
\begin{eqnarray}
G_{\mbox{\scriptsize step+delta}}(L) = \left[1-\left(\frac{d+1}{d+2}\right)\left(\frac{D}{L}\right)\right]\Theta(L-D),
\end{eqnarray}
which converges to $G_{\mbox{\scriptsize step}}(L)$ as $d\rightarrow +\infty$.  This is precisely the 
statement that ``effective'' pair correlations in the step+delta-function process asymptotically
vanish in high dimensions, which is the decorrelation principle as we have presented it 
for periodic structures.  

We can also show that the cumulative coordination number $Z(R)$ [c.f. \eref{ZR}], 
a particular smoothing operation on the pair correlation function, can be insensitive 
to decorrelation in disordered point patterns and is therefore not the appropriate quantity to 
examine for periodic structures.  Direct 
calculation of $Z(R)$ for the 
$g_2$-invariant processes above shows that the cumulative coordination numbers 
for both systems are identical and 
are given by
\begin{eqnarray}
Z(R) = \left[\rho v(R)-1\right]\Theta(R-D),
\end{eqnarray}
which is equivalent to the hyperuniformity condition $Z(R) + 1 = \rho v(R)$ as $R\rightarrow+\infty$.
Therefore, although $Z(R)$ can distinguish hyperuniform and non-hyperuniform point patterns, 
it provides little information about the behavior of asymptotic correlations in high dimensions.


\begin{thebibliography}{}
\bibitem{FNPS}
The Poisson summation formula generalizes the Fourier series of an appropriately 
well-behaved function $f$ that is periodic with respect the fundamental cell of a Bravais lattice 
$\Lambda$.  Namely, for every vector $\mathbf{v} \in \mathbb{R}^d$, 
$\sum_{\mathbf{p} \in \Lambda} f(\mathbf{p}+\mathbf{v}) = [v_F(\Lambda)]^{-1}
\sum_{\mathbf{q} \in \Lambda^*} \exp(-2\pi i \mathbf{q}\cdot \mathbf{v}) \hat{f}(\mathbf{q})$,
where $v_F(\Lambda)$ is the volume of the fundamental cell for $\Lambda$ and
$\hat{f}$ is the unitary Fourier transform of the function $f$. See Ref. [42]
for an application of Poisson summation to maximally dense sphere packings.
\bibitem{FNAd}
Using these properties, one can recursively write down the basis vectors $\mathbf{e}_j$ for 
the $A_d$ lattice.  For example, assuming $a = 1$, we have in two dimensions $\mathbf{e}_1 = (1, 0)^{\mbox{\scriptsize T}}$
and $\mathbf{e}_2 = (1/2, \sqrt{3}/2)^{\mbox{\scriptsize T}}$.  The basis vectors for $d = 3$ are
$\mathbf{e}_1 = (1, 0, 0)^{\mbox{\scriptsize T}}$, $\mathbf{e}_2 = (1/2, \sqrt{3}/2, 0)^{\mbox{\scriptsize T}}$, 
and $\mathbf{e}_3 = (1/2, \sqrt{3}/6, \sqrt{6}/3)^{\mbox{\scriptsize T}}$.  The $d = 4$ vectors are
$\mathbf{e}_1 = (1,0,0,0)^{\mbox{\scriptsize T}}$, $\mathbf{e}_2 = (1/2,\sqrt{3}/2,0,0)^{\mbox{\scriptsize T}}$,
$\mathbf{e}_3 = (1/2,\sqrt{3}/6,\sqrt{6}/3,0)^{\mbox{\scriptsize T}}$, and 
$\mathbf{e}_4 = (1/2,\sqrt{3}/6,\sqrt{6}/12,\sqrt{10}/4)^{\mbox{\scriptsize T}}$.
\bibitem{FN4}
Note that $E_V$ for a Poisson point pattern decays exponentially as $E_V(R) = \exp[-\rho v(R)]$,
implying that this structure possesses an infinite covering radius.  However, the long tail of
this function
does not contribute substantially enough for the quantizer error to be larger than the kagom\'e lattice
in sufficiently high dimensions.
\bibitem{FNiso}
Isostaticity, where $Z = 2d_f$ with $d_f$ equal to the number of degrees of freedom per particle,
has been closely linked to the notions of ``generic'' or ``random'' jammed sphere packings.    
However, for packings with high symmetry, including the hypercubic $\mathbb{Z}^d$ 
and kagom\'e crystals, the isostatic condition does not imply collective or strict jamming.
See Ref. [13]
for a more complete discussion of this fundamentally important point.
\end{thebibliography}
\end{document}